\documentclass[10pt]{report}

\usepackage[utf8]{inputenc}
\DeclareUnicodeCharacter{2212}{-}
\usepackage{amsmath}
\usepackage{amssymb}
\usepackage{amsfonts}
\usepackage{pifont}
\usepackage{pdfpages}
\usepackage[most]{tcolorbox} 
\usepackage{array}

\usepackage[backend=biber,
style=numeric,
sortcites,
natbib=true,
sorting=none]{biblatex}
\usepackage[]{titlesec}
\titleformat{\chapter}[display]
  {\normalfont\bfseries}{}{0pt}{\LARGE}
  \titlespacing*{\chapter}{0pt}{-50pt}{10pt}

\usepackage[T1]{fontenc}    
\usepackage{enumitem}

\usepackage{amsfonts}       
\usepackage{nicefrac}       
\usepackage{microtype}      
\usepackage{listings}
\usepackage{dashrule}
\usepackage{wrapfig}
\usepackage{mdframed}

\usepackage[most]{tcolorbox}
\usepackage{tabularray}

\usepackage{amsmath,amsfonts,bm}

\def\eqref#1{equation~\ref{#1}}

\def\1{\bm{1}}

\DeclareMathAlphabet{\mathsfit}{\encodingdefault}{\sfdefault}{m}{sl}
\SetMathAlphabet{\mathsfit}{bold}{\encodingdefault}{\sfdefault}{bx}{n}

\usepackage{amsfonts}

\usepackage[breaklinks=true,colorlinks,citecolor=black,bookmarks=false]{hyperref}
\hypersetup{
    colorlinks=false,
	linkcolor=blue,
	filecolor=magenta,      
	urlcolor=blue,
	citecolor=black,
	pdfinfo={
		Title={Superintelligence Strategy: Expert Version},
		Author={Dan Hendrycks, Eric Schmidt, Alexandr Wang},
		Subject={National Security, AI Safety},
		Keywords={geopolitics, ai safeguards, superintelligence, compute, deterrence}
	}
}

\usepackage{times}
\usepackage{url}
\usepackage{lipsum}
\usepackage{bbm}
\usepackage{wrapfig}
\usepackage{graphicx}
\usepackage{subcaption}
\usepackage{multirow}
\usepackage{todonotes}
\usepackage{amssymb}
\usepackage{amsfonts}
\usepackage{microtype}      
\usepackage{cleveref}
\usepackage{titling} 
\usepackage{svg}
\usepackage[most]{tcolorbox}
\usepackage{tabularray}

\usepackage{spverbatim}

\usepackage{float}
\newfloat{storyfloat}{htbp}{loa}
\floatname{storyfloat}{Story}

\newcounter{story}[section]

\newcounter{vision}[section]

\usepackage{arydshln}

\usepackage{booktabs}       
\usepackage{tabularray}
\UseTblrLibrary{booktabs} 
\UseTblrLibrary{amsmath}  
\definecolor{dodgerblue}{RGB}{53, 133, 212}
\definecolor{limegreen}{RGB}{79, 171, 79}
\usepackage{dashrule}       

\title{
\vspace{-25pt}
\hrule height 4pt
\vskip 0.25in
{\LARGE\bf Superintelligence Strategy: Expert Version}
\vskip 0.29in
\hrule height 1pt
\vskip 0.09in
}
\date{}

\renewenvironment{abstract}%
{%
  \centerline%
  {\large\bf Abstract}%
  \begin{quote}%
}
{
  \par%
  \end{quote}%
  \vskip 1ex%
}

\author{\textbf{Dan Hendrycks} \quad \textbf{Eric Schmidt} \quad \textbf{Alexandr Wang} 
}

\newcommand{\reviewer}[3]{
	\expandafter\newcommand\csname #1\endcsname[1]{
		\textcolor{#3}{[#2: ##1]}
	}
}
\definecolor{neonpurple}{rgb}{0.3,0,1}
\reviewer{dan}{Dan}{neonpurple}

\AtBeginBibliography{\small}

\usepackage{etoolbox}

\makeatletter
\patchcmd{\chapter}
  {\if@openright\cleardoublepage\else\clearpage\fi}
  {\vspace*{5\baselineskip}}
  {}{}
\makeatother

\begin{document}
\includepdf[pages={1}]{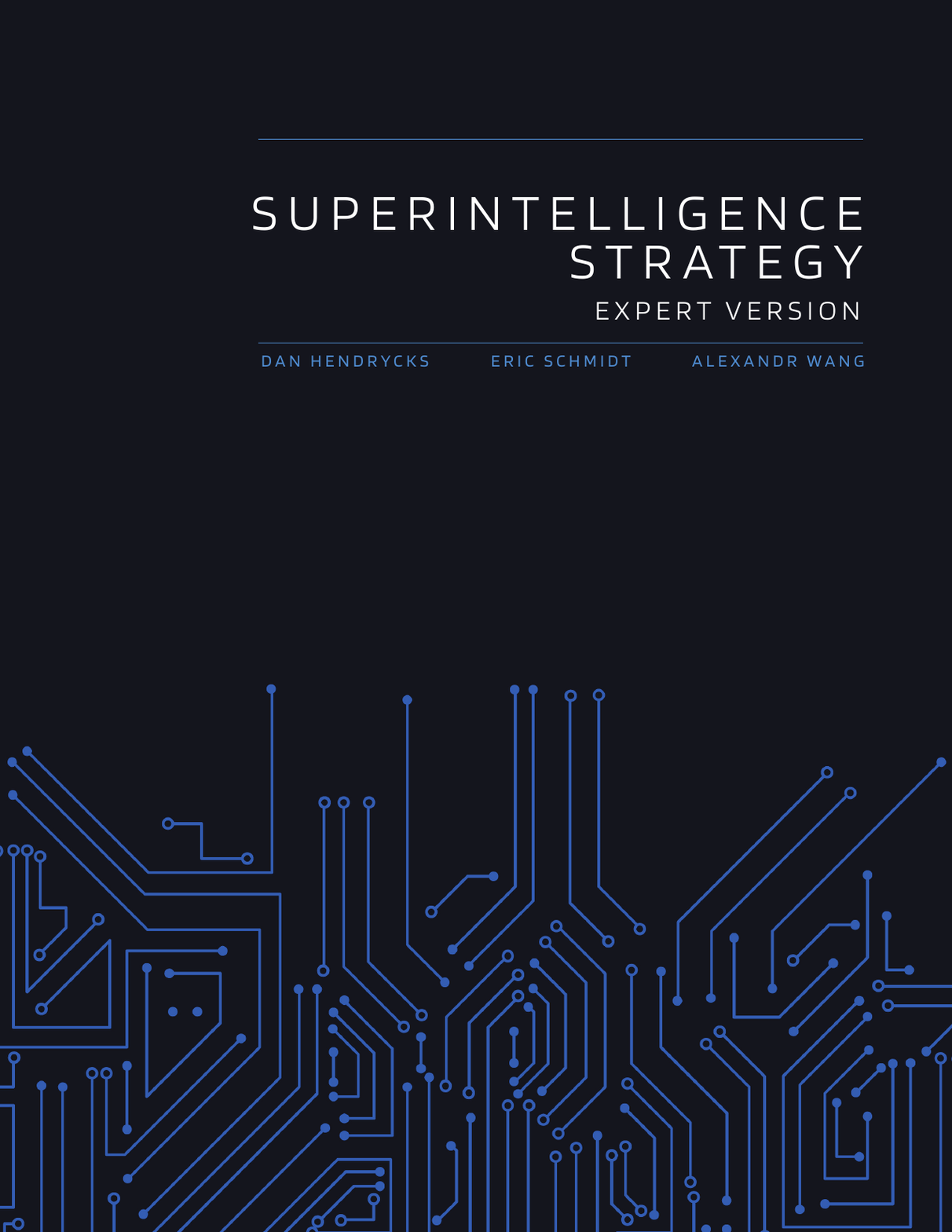}


{
  \let\newpage\relax   
  \let\clearpage\relax 
  \maketitle
  \vspace*{-160pt}

  \begin{abstract}
      \normalsize
    Rapid advances in AI are beginning to reshape national security. Destabilizing AI developments could rupture the balance of power and raise the odds of great-power conflict, while widespread proliferation of capable AI hackers and virologists would lower barriers for rogue actors to cause catastrophe. Superintelligence---AI vastly better than humans at nearly all cognitive tasks---is now anticipated by AI researchers. Just as nations once developed nuclear strategies to secure their survival, we now need a coherent superintelligence strategy to navigate a new period of transformative change. We introduce the concept of Mutual Assured AI Malfunction (MAIM): a \textit{deterrence} regime resembling nuclear mutual assured destruction (MAD) where any state’s aggressive bid for unilateral AI dominance is met with preventive sabotage by rivals. Given the relative ease of sabotaging a destabilizing AI project---through interventions ranging from covert cyberattacks to potential kinetic strikes on datacenters---MAIM already describes the strategic picture AI superpowers find themselves in. Alongside this, states can increase their \textit{competitiveness} by bolstering their economies and militaries through AI, and they can engage in \textit{nonproliferation} to rogue actors to keep weaponizable AI capabilities out of their hands. Taken together, the three-part framework of deterrence, nonproliferation, and competitiveness outlines a robust strategy to superintelligence in the years ahead.


  \end{abstract}
}

\newpage

\vspace*{-6\baselineskip}
\chapter{Executive Summary}


Rapid advances in AI are poised to reshape nearly every aspect of society. Governments see in these dual-use AI systems a means to military dominance, stoking a bitter race to maximize AI capabilities. Voluntary industry pauses or attempts to exclude government involvement cannot change this reality. These systems that can streamline research and bolster economic output can also be turned to destructive ends, enabling rogue actors to engineer bioweapons and hack critical infrastructure. ``Superintelligent'' AI surpassing humans in nearly every domain would amount to the most precarious technological development since the nuclear bomb. Given the stakes, superintelligence is inescapably a matter of national security, and an effective superintelligence strategy should draw from a long history of national security policy.

\subsection*{Deterrence}
A race for AI-enabled dominance endangers all states. If, in a hurried bid for superiority, one state inadvertently loses control of its AI, it jeopardizes the security of all states. Alternatively, if the same state succeeds in producing and controlling a highly capable AI, it likewise poses a direct threat to the survival of its peers. In either event, states seeking to secure their own survival may threaten to sabotage destabilizing AI projects for deterrence. A state could try to disrupt such an AI project with interventions ranging from covert operations that degrade training runs to physical damage that disables AI infrastructure. Thus, we are already approaching a dynamic similar to nuclear Mutual Assured Destruction (MAD), in which no power dares attempt an outright grab for strategic monopoly, as any such effort would invite a debilitating response. This strategic condition, which we refer to as \textbf{Mutual Assured AI Malfunction (MAIM)}, represents a potentially stable deterrence regime, but maintaining it could require care. We outline measures to maintain the conditions for MAIM, including clearly communicated escalation ladders, placement of AI infrastructure far from population centers, transparency into datacenters, and more.\looseness=-1

\subsection*{Nonproliferation}
While deterrence through MAIM constrains the intent of superpowers, all nations have an interest in limiting the AI capabilities of terrorists. Drawing on nonproliferation precedents for weapons of mass destruction (WMDs), we outline three levers for achieving this. Mirroring measures to restrict key inputs to WMDs such as fissile material and chemical weapons precursors, \textit{compute security} involves knowing reliably where high-end AI chips are and stemming smuggling to rogue actors. Monitoring shipments, tracking chip inventories, and employing security features like geolocation can help states account for them. States must prioritize \textit{information security} to protect the model weights underlying the most advanced AI systems from falling into the hands of rogue actors, similar to controls on other sensitive information. Finally, akin to screening protocols for DNA synthesis services to detect and refuse orders for known pathogens, AI companies can be incentivized to implement technical \textit{AI security} measures that detect and prevent malicious use.

\subsection*{Competitiveness}
Beyond securing their survival, states will have an interest in harnessing AI to bolster their competitiveness, as successful AI adoption will be a determining factor in national strength. Adopting AI-enabled weapons and carefully integrating AI into command and control is increasingly essential for \emph{military} strength. Recognizing that \emph{economic} security is crucial for national security, domestic capacity for manufacturing high-end AI chips will ensure a resilient supply and sidestep geopolitical risks in Taiwan. Robust \emph{legal frameworks} governing AI agents can set basic constraints on their behavior that follow the spirit of existing law. Finally, governments can maintain \emph{political stability} through measures that improve the quality of decision-making and combat the disruptive effects of rapid automation. \\

By detecting and deterring destabilizing AI projects through intelligence operations and targeted disruption, restricting access to AI chips and capabilities for malicious actors through strict controls, and guaranteeing a stable AI supply chain by investing in domestic chip manufacturing, states can safeguard their security while opening the door to unprecedented prosperity.

\newpage
\tableofcontents
\newpage

\vspace*{-3\baselineskip}
\chapter{Introduction}

Artificial Intelligence (AI) is rapidly transforming multiple facets of society, with advances arriving at a pace and scale that few anticipated. These developments compel policymakers, technologists, and strategists to address a widening spectrum of issues, from economic shifts driven by automation to strategic concerns about global competition. As with any transformative technology, AI presents both significant opportunities and formidable risks.

Among these challenges, the dual-use nature of AI---its capacity for both civilian and military applications---emerges as a critical factor. Unlike specialized technological tools, AI spans virtually every sector, including finance, healthcare, and defense. This broad applicability, coupled with its rapid evolution, creates a risk landscape that is expansive and difficult to predict. Strategic actors must contend with potential misuse, risks of geopolitical escalation, and the need for frameworks to govern systems whose capabilities may surpass human oversight.

To navigate these complexities, many have turned to analogies. AI has been compared to electricity for its general-purpose nature, to traditional software for its economic importance, or to the printing press for its cultural impact. While these comparisons provide useful entry points, they fail to emphasize the grave national security implications of AI. A more productive analogy lies between AI and catastrophic dual-use nuclear, chemical, and biological technologies. Like them, AI will be integral to a nation’s power while posing the potential for mass destruction. A brief examination of the historical parallels between AI and the nuclear age can highlight the gravity of our current situation.

In 1933, the leading scientist Ernest Rutherford dismissed the notion of harnessing atomic power as ``moonshine.'' The very next day, Leo Szilard read Rutherford’s remarks and sketched the idea of a nuclear chain reaction that ultimately birthed the nuclear age. Eventually figures such as J.\ Robert Oppenheimer recognized the dual nature of their work. Today, AI is at a similar stage. Previously considered science fiction, AI has advanced to the point where machines can learn, adapt, and potentially exceed human intelligence in certain areas. AI experts including Geoffrey Hinton and Yoshua Bengio, pioneers in deep learning, have expressed existential concerns about the technologies they helped create \cite{cais2023}.

As AI's capabilities are becoming more evident, nations and corporations are heavily investing to gain a strategic advantage. The Manhattan Project, which consumed 0.4\% of the U.S.\ GDP, was driven by the need to develop nuclear capabilities ahead of others. Currently, a similar urgency is evident in the global effort to lead in AI, with investment in AI training doubling every year for nearly the past decade. Several ``AI Manhattan Projects'' aiming to eventually build superintelligence are already underway, financed by many of the most powerful corporations in the world.

\begin{figure}[h]
    \centering
    \includegraphics[width=\textwidth]{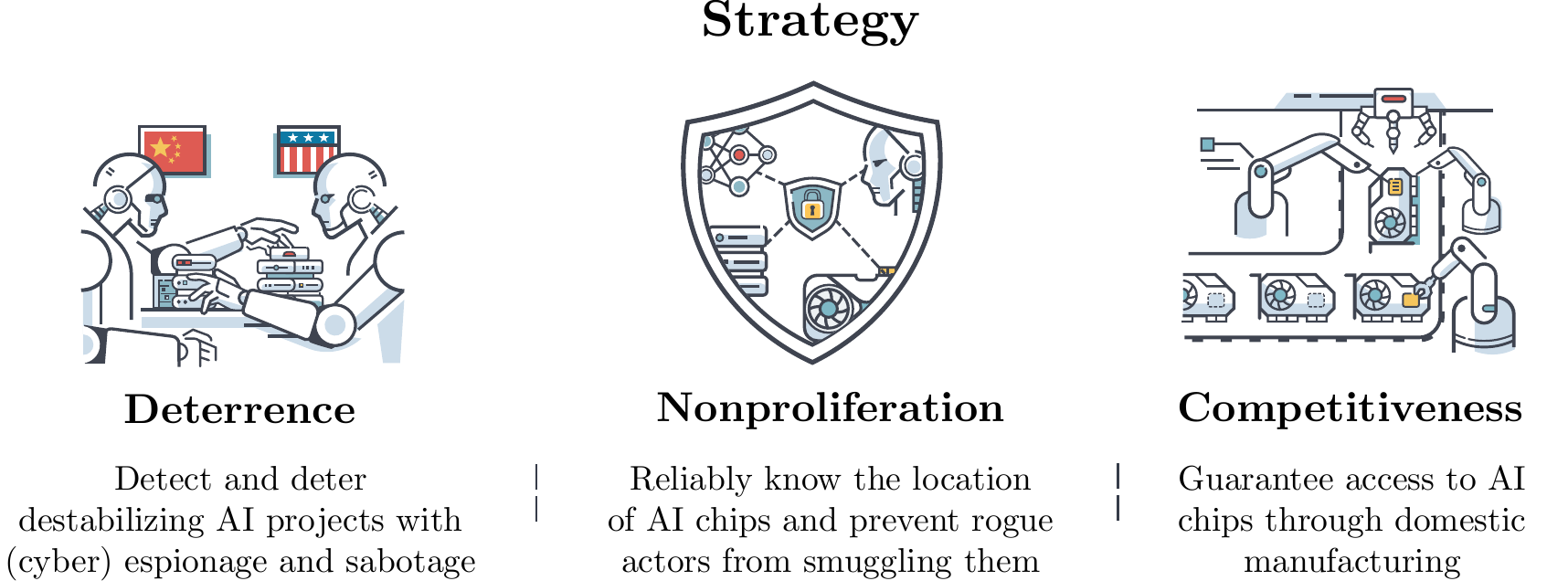}
    \caption{Effective strategies for managing advanced AI can draw from national security precedents in handling previous potentially catastrophic dual-use technology.}
    \label{fig:strategy}
\end{figure}

However, the rapid advancement of AI technologies introduces significant uncertainties for international stability and security. The introduction of nuclear weapons altered international relations, granting influence to those who possessed them and leading to an arms race. The Cuban Missile Crisis highlighted how close the world came to nuclear war. Nuclear annihilation has been avoided thus far despite significant tensions between nuclear states in part through the deterrence principle of Mutual Assured Destruction (MAD), where any nuclear use would provoke an in-kind response. In the AI era, a parallel form of deterrence could emerge---what might be termed ``Mutual Assured AI Malfunction'' (MAIM)---where states' AI projects are constrained by mutual threats of sabotage.

\begin{wrapfigure}{r}{0.5\textwidth}
    \vspace{-5pt}
    \centering
    \includegraphics[width=\linewidth]{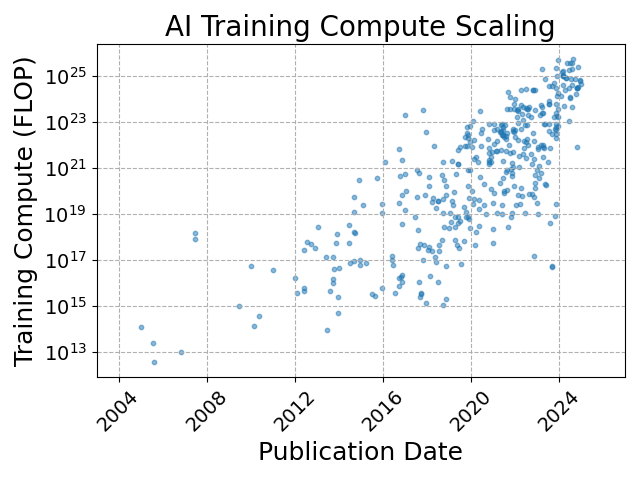}
    \caption{The amount of compute used to create AI models has been increasing exponentially for decades \cite{epoch2023aitrends}.\looseness=-1}
    \vspace{-10pt}
    \label{fig:computetime}
\end{wrapfigure}

The risks are not limited to state competition; advanced dual-use technologies can also be exploited by non-state actors. Just as the spread of nuclear capabilities raised concerns about misuse, the availability of AI systems presents new challenges. Malicious actors could use AI to develop weapons of mass destruction or conduct large-scale cyberattacks on critical infrastructure. The accessibility of unsecured or open-weight AI increases these risks, highlighting the need for careful policies and safeguards.

In the nuclear era, uranium became the linchpin of atomic power. States that secured it could enforce regulations, negotiate treaties, and limit the spread of destructive capabilities. In the realm of AI, computing resources---especially AI chips---have a similar strategic weight, fueling rivalries and shaping geopolitical calculations. This dynamic is evident in places such as Taiwan, central to AI chip production, where rising tensions could have extensive consequences. Nations have a shared interest in controlling access to AI chips to keep them out of the hands of ``rogue actors''---terrorist groups and small pariah states---echoing the logic once applied to uranium.

Despite these challenges, AI offers significant opportunities. Nuclear technology, while introducing the threat of mass destruction, also provided a new energy source that transformed societies. AI has the potential to drive advancements across various sectors, from medical breakthroughs to economic automation. Embracing AI's benefits is important for economic growth and progress in the modern world.

The challenges AI poses are far too broad, and far too serious, for piecemeal measures. What is needed is a comprehensive strategy, one that does not shy from the unsettling implications of advanced AI. As with Herman Kahn's famous analysis of nuclear strategy \cite{kahn1960thermonuclear}, superintelligence strategy requires ``thinking about the unthinkable.'' In this paper, we propose such a strategy and grapple with these fundamental questions along the way: What should be done about lethal autonomous weapons? Catastrophic malicious use? Powerful open-weight AIs? AI-powered mass surveillance? How can society maintain a shared grasp of reality? What should be done about AI rights? How can humans maintain their status in a world of mass automation?

We argue that the most effective framework for addressing AI’s challenges is to view it through a national security lens. Drawing on lessons from previous dual-use technologies while tailoring them to the distinct demands of AI can help safeguard against catastrophic misuse, maintain geopolitical stability, and ensure that the broader Western world remains at the forefront. 



\begin{figure}[h]
\hspace{-20pt} 
\begin{booktabs}{
  colspec={p{3.5cm} p{4.5cm} p{7.2cm}}, 
  row{odd}={blue9},                    
  row{1}={white}                       
}
\toprule
\textbf{Topic} & \textbf{Tame Technical Subproblem} & \textbf{Wicked Problems} \\
\midrule
Malicious Use 
& \begin{tabular}{@{}l@{}}
Train AI systems to \\
refuse harmful requests
\end{tabular}
& \begin{tabular}{@{}l@{}}
Coordinate nonproliferation among countries,\\
incentivize AI deployers to prevent misuse, \\
validate know-your-customer information
\end{tabular}
\\

Deterrence
& \begin{tabular}{@{}l@{}}
Prepare cyberattacks for \\
AI datacenters
\end{tabular}
& \begin{tabular}{@{}l@{}}
Define escalation thresholds and signal intent, \\
surveil rival AI projects, \\
adapt verification measures as AI evolves
\end{tabular}
\\

Compute Security
& \begin{tabular}{@{}l@{}}
Upgrade AI chip firmware to \\
add geolocation functionality
\end{tabular}
& \begin{tabular}{@{}l@{}}
Enforce export controls to track AI chips, \\
report suspicious shipments to allies, \\
verify the decommissioning of obsolete AI chips
\end{tabular}
\\

Information Security
& \begin{tabular}{@{}l@{}}
Patch known vulnerabilities in \\
AI developers’ computer systems
\end{tabular}
& \begin{tabular}{@{}l@{}}
Counter insider threats, balance security measures \\
with productivity, prevent ideological leaks
\end{tabular}
\\

Military Strength
& \begin{tabular}{@{}l@{}}
Design military drones
\end{tabular}
& \begin{tabular}{@{}l@{}}
Secure supply chains, adapt military \\
organizational structures to diffuse AI, \\
maintain meaningful human control over \\
AI-augmented command and control
\end{tabular}
\\

Economic Strength
& \begin{tabular}{@{}l@{}}
Improve AI performance on \\
economically valuable tasks
\end{tabular}
& \begin{tabular}{@{}l@{}}
Draw from legal frameworks for AI agents, \\
establish and evolve multiagent infrastructure, \\
preserve stability amid automation
\end{tabular}
\\

Loss of Control
& \begin{tabular}{@{}l@{}}
Research methods to make \\
current AIs follow instructions
\end{tabular}
& \begin{tabular}{@{}l@{}}
Steer a population of rapidly evolving AIs \\
during an intelligence recursion, \\
co-evolve safeguards with AI systems, \\
red team AIs to identify unknown unknowns
\end{tabular}
\\
\bottomrule
\end{booktabs}
\caption{AI risk management contains multiple wicked problems and is not primarily a technical challenge. Tame \textit{technical problems} have well-defined boundaries and criteria for success, lending themselves to systematic experimentation. By contrast, \textit{wicked problems} \cite{Rittel1973} are open-ended, carry ambiguous requirements, and often produce unintended consequences. They demand ongoing adaptation rather than purely technical fixes, since each attempt at a solution can give rise to new difficulties.}
\label{fig:wicked-problems}
\end{figure}

\chapter{AI Is Pivotal for National Security}

AI holds the potential to reshape the balance of power. In the hands of state actors, it can lead to disruptive military capabilities and transform economic competition. At the same time, terrorists may exploit its dual-use nature to orchestrate attacks once within the exclusive domain of great powers. It could also slip free of human oversight.

This chapter examines these three threats to national security: rival states, rogue actors, and uncontrolled AI systems. It closes by assessing why existing strategies fall short of managing these intertwined threats.

\begin{figure}[t]
    \centering
    \includegraphics[width=\textwidth]{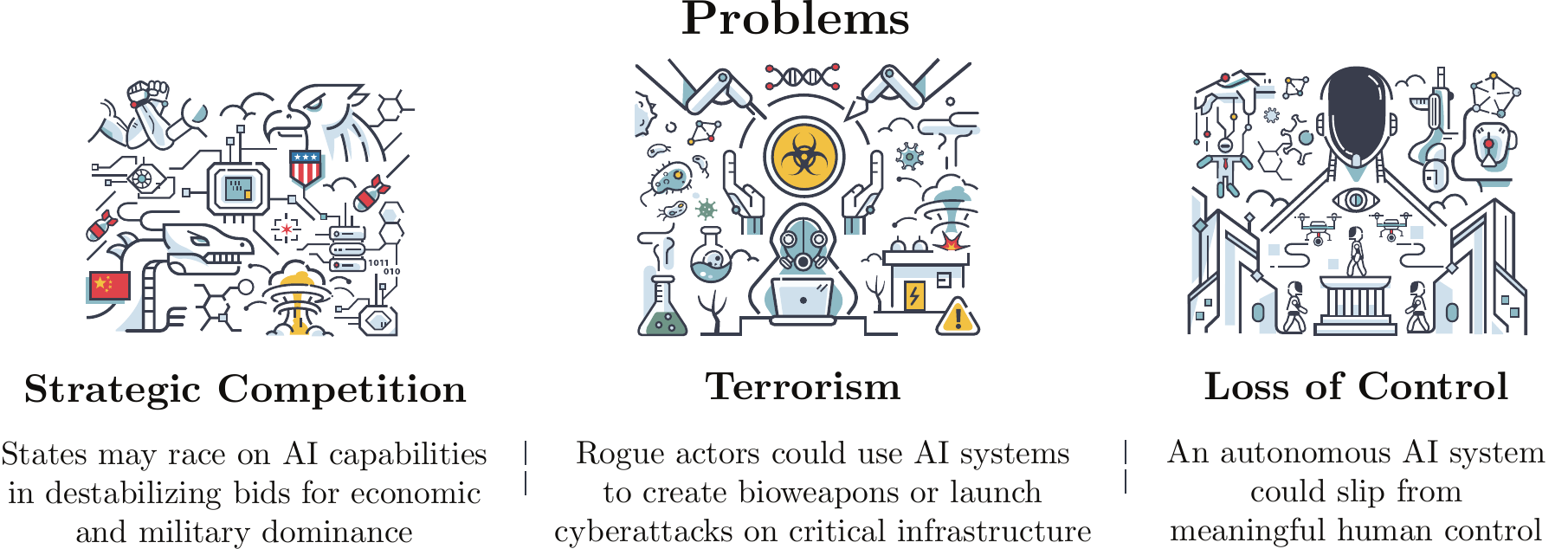}
    \caption{States, terrorists, and AIs are threats to national security.}
    \label{fig:threats}
\end{figure}

\section{Strategic Competition}

In an international system with no central authority, states prioritize their own strength to ensure their own security. This competition arises not from a desire for dominance but from the necessity of safeguarding national interests. They exist in an environment where threats can emerge unexpectedly and assistance from others is uncertain. Some states may rise in their power and provoke alarm in their rivals, a pattern called the Thucydides Trap \cite{allison2017destined}. Consequently, states seek to preserve their relative power.

In this environment, the impact of AI on state power looms large. AI may transform the foundations of economic and military power. Its ability to automate labor could become the source of economic competitiveness. In the military sphere, it could be used to dominate rivals. We begin by looking at economic power, then turn to its greatest military implications.

\subsection{Shifting Basis of Economic Power}
\paragraph{AI Chips as the Currency of Economic Power.} As AI becomes more and more integrated in the economy, the possession of advanced AI chips may define a nation's power. Historically, wealth and population size underpinned a state's influence; however, the automation of tasks through AI alters this dynamic. A collection of highly capable AI agents, operating tirelessly and efficiently, rivals a skilled workforce, effectively turning capital into labor. In this new paradigm, power will depend both on the capability of AI systems and the number of AI chips on which they could run. Nations with greater access to AI chips could outcompete others economically.

\subsection{Destabilization Through Superweapons}

States have long pursued weapons that could confer a decisive advantage over rivals. AI systems introduce new avenues for such pursuit, raising questions about whether certain breakthroughs could undermine deterrence and reorder global power structures.

\paragraph{AI Could Enable Military Dominance.} Advanced AI systems may drive technological breakthroughs that alter the strategic balance, similar to the introduction of nuclear weapons, and could generate strategic surprise that catches rivals off-guard \cite{wonderweapons}. Such a ``superweapon'' may grant two tiers of advantages. One, which might be called ``subnuclear dominance,'' would allow a state to project power widely and subdue adversaries without disrupting nuclear deterrence. The second possibility---a ``strategic monopoly'' on power---would upend the nuclear balance entirely and could establish one state's complete dominance and control, leaving the fate of rivals subject to its will.

\paragraph{Possible Superweapons.}
Subnuclear superweapons---such as an AI-enabled cyberweapon that can suddenly and comprehensively destroy a state's critical infrastructure, exotic EMP devices, and next-generation drones---could confer sweeping advantages without nullifying an adversary's nuclear deterrent. Some superweapons might erode mutual assured destruction outright. A ``transparent ocean'' would threaten submarine stealth, revealing the location of nuclear submarines. AIs might be able to pinpoint all hardened mobile nuclear launchers, further undermining the nuclear triad. AIs could undermine situational awareness and sow confusion by generating elaborate deceptions---a ``fog of war machine'' \cite{geist2023fog}---that mask true intentions or capabilities. A defensive superweapon possibility is an anti-ballistic missile system that eliminates an adversary’s ability to strike back. Lastly, some superweapons remain beyond today’s foresight---``unknown unknowns'' that could undermine strategic stability.

\paragraph{Implications of Superweapons.} Superintelligence is not merely a new weapon, but a way to fast-track all future military innovation. A nation with sole possession of superintelligence might be as overwhelming as the Conquistadors were to the Aztecs. If a state achieves a strategic monopoly through AI, it could reshape world affairs on its own terms. An AI-driven surveillance apparatus might enable an \emph{unshakable totalitarian regime}, transforming governance at home and leverage abroad.

The mere pursuit of such a breakthrough could, however, tempt rivals to act before their window closes. Fear that another state might soon rapidly grow in power has led observers to contemplate measures once seen as unthinkable. In the nuclear era, Bertrand Russell, ordinarily a staunch pacifist, proposed preventive nuclear strikes on the Soviet Union to thwart its rise \cite{Perkins1995}, while the United States seriously pondered crippling the Chinese nuclear program during the early 60s \cite{Burr2000}. Faced with the specter of superweapons and an AI-enabled strategic monopoly on power, some leaders may turn to \emph{preventive action} \cite{fearon1995rationalist}. Rather than only relying on cooperation or seeking to outpace their adversaries, they may consider sabotage or datacenter attacks, if the alternative is to accept a future in which one’s national survival is perpetually at risk.


Superweapons and shifting economic power can redefine strategic competition. To grasp the full magnitude of AI's impacts on national security, we turn next to rogue actors, and later to AI systems that slip from human control.

\section{Terrorism}

\paragraph{AI's Dual-Use Capabilities Amplify Terrorism Risks.}
As AI capabilities increase, it will likely be important not only in the context of state-level competition but also as an amplifier of terrorist capabilities. Technologies that can revolutionize healthcare or simplify software development also have the potential to empower individuals to create bioweapons and conduct cyberattacks. This amplification effect lowers the barriers for terrorists, enabling them to execute large-scale attacks that were previously limited to nation-states. This section examines two critical areas where AI intensifies terrorism risks: lowering barriers to bioweapon development, and lowering barriers to cyberattacks against critical infrastructure.

\subsection{Bioterrorism}
\paragraph{AI Lowers Barriers to Bioterrorism.}
Consider Aum Shinrikyo, the Japanese cult that orchestrated the 1995 Tokyo subway sarin attack. Operating with limited expertise, they managed to produce and deploy a chemical weapon in the heart of Tokyo's transit system, killing 13 people and injuring over 5,000. The attack paralyzed the city, instilling widespread fear and demonstrating the havoc that determined non-state actors can wreak.

With AI assistance, similar groups could achieve far more devastating results. AI could provide step-by-step guidance on designing lethal pathogens, sourcing materials, and optimizing methods of dispersal. What once required specialized knowledge and resources could become accessible to individuals with malevolent intent, dramatically increasing the potential for catastrophic outcomes. Indeed, some cutting-edge AI systems without bioweapons safeguards already exceed expert-level performance on numerous virology benchmarks \cite{gotting2025vct}.


Engineered pathogens could surpass historical pandemics in scale and lethality. The Black Death killed in the vicinity of half of Europe's population without human engineering. Modern bioweapons, enhanced by AI-driven design, could exploit vulnerabilities in human biology with unprecedented precision, creating contagions that evade detection and resist treatment. While most discussions of bioweapons are secret, some scientists have openly warned of ``mirror bacteria,'' engineered with reversed molecular structures that could evade the immune defenses that normally keep pathogens at bay. Though formidable to create, they have prompted urgent appeals from leading researchers to halt development \cite{mirrorbacteria}, lest they unleash a catastrophe unlike any our biosphere has known. In contrast to other weapons of mass destruction, biological agents can self-replicate, allowing a small initial release to spiral into a worldwide calamity.


\subsection{Cyberattacks on Critical Infrastructure}
\paragraph{AI Lowers Barriers to Cyberattacks Against Critical Infrastructure.}
Our critical infrastructure, including power grids and water systems, is more fragile than it may appear. A hack targeting digital thermostats could force them to cycle on and off every few minutes, creating damaging power surges that burn out transformers---critical components that can take years to replace. Another approach would be to exploit vulnerabilities in Supervisory Control and Data Acquisition software, compelling sudden load shifts and driving transformers beyond safe limits. At water treatment facilities, tampered sensor readings could fail to detect a dangerous mixture, and filtration processes could be halted at key intervals, allowing contaminants to enter the municipal supply undetected---all without needing on-site sabotage. The Department of Homeland Security has cautioned that AI could be employed by malicious actors to exploit vulnerabilities in these systems \cite{dhs2024ai_framework}. Presently, only highly skilled operatives or nation-states possess the expertise to conduct such sophisticated operations, like the Stuxnet worm that damaged Iran’s nuclear facilities. However, AI could democratize this capability, providing rogue actors with tools to design and execute attacks with greater accessibility, speed, and scale.\looseness=-1

AI-driven programs could tirelessly scan for vulnerabilities, adapt to defensive measures, and coordinate assaults across multiple targets simultaneously. The automation of complex hacking tasks reduces the need for specialized human expertise. This shift could enable individuals to cause disruptions previously achievable only by governments. Moreover, AI-assisted attacks may be harder to trace back to a specific attacker. This could incentivize adversaries to conduct such an attack, if they believe their target will not be able to determine they were the perpetrator. The difficulty of attribution complicates responses and heightens the risk of escalation between great powers, potentially leading to severe conflicts.

\subsection{Offense-Defense Balance}

\begin{wrapfigure}{r}{0.5\textwidth}
    \centering
    \includegraphics[width=\linewidth]{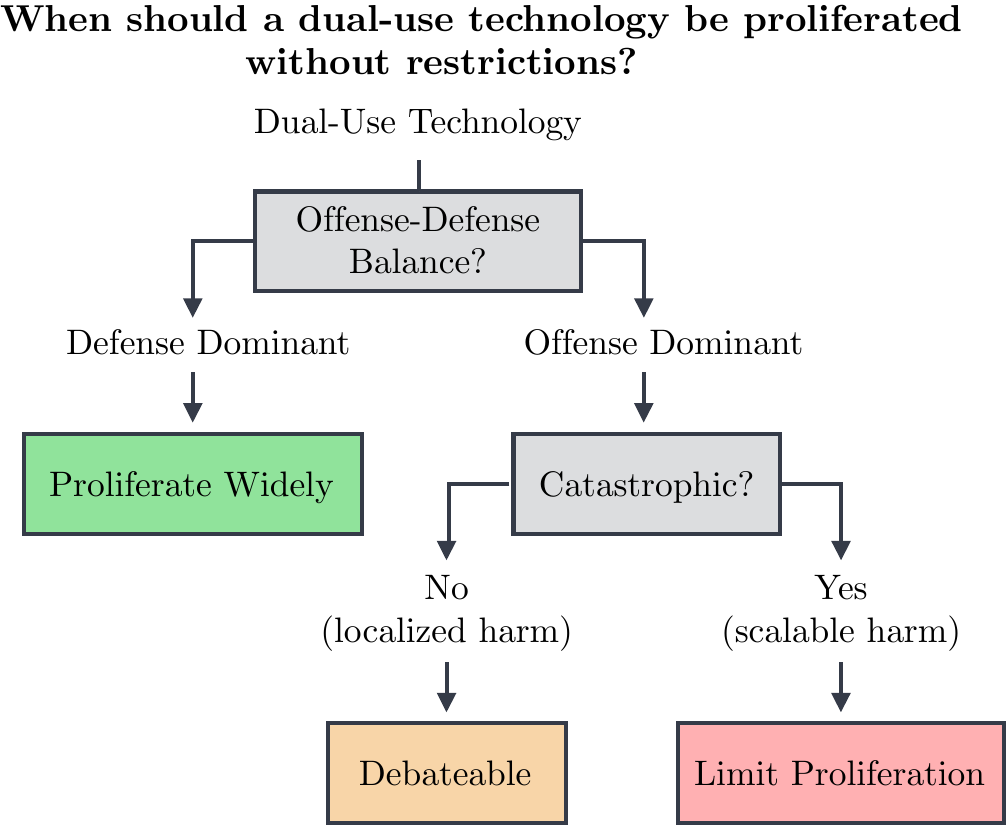}
    \caption{Defense-dominant dual-use technology should be widely proliferated, while catastrophic offense-dominant dual-use technology should not.\looseness=-1}
    \label{fig:proliferationflowchart}
\end{wrapfigure}

\paragraph{AI Is Often Offense Dominant.}
Some argue that broad access to AI technologies could strengthen defensive capabilities. However, in both biological and critical infrastructure contexts, attackers hold significant advantages. In biotechnology, developing cures or defenses against engineered pathogens is complex and time-consuming, which would lag behind the creation and deployment of new threats. The rapid self-replication of biological agents amplifies damage before effective countermeasures can be implemented. Many viruses still do not have a cure.

Critical infrastructure systems often suffer from ``patch lag,'' resulting in software remaining unpatched for extended periods, sometimes years or decades. In many cases, patches cannot be applied in a timely manner because systems must operate without interruption, the software remains outdated because its developer went out of business, or interoperability constraints require specific legacy software. Adversaries have enduring opportunities to exploit vulnerabilities within critical infrastructure. As AI tools advance, even novice adversaries could automate the discovery of software vulnerabilities, coordinating attacks at scale. While software that receives frequent updates, such as Chrome, does not suffer from substantial patch lag and can be more resilient, critical infrastructure remains at a distinct disadvantage. Under these conditions, an adversary needs to find only one overlooked vulnerability, while defenders grapple with the far more daunting task of handling every corner and patching every vulnerability if they hope to achieve defense dominance \cite{hendrycks2022unsolvedproblemsmlsafety, newman2024cybersecurity}.

Historical efforts to shift the offense-defense balance illustrate inherent challenges with WMDs. During the Cold War, the Strategic Defense Initiative aimed to develop systems to intercept incoming nuclear missiles and render nuclear arsenals obsolete. Despite significant investment, creating an impermeable defense proved unfeasible, and offense remained dominant.


\section{Loss of Control}

We now shift from threats involving rival states and terrorists to a new source of threat: the possibility of losing control over an AI system itself.
Here, AIs do not just amplify existing threats but create new paths to mass destruction. A loss of control can occur if militaries and companies grow so dependent on automation that humans no longer have meaningful control, if an individual deliberately unleashes a powerful system, or if automated AI research outruns its development safeguards. While this threat is the least understood, its severity can be great enough to permanently undermine national security.

\subsection{Erosion of Control}
    
Waves of automation, once incremental, may strike entire sectors at once and leave human workers abruptly displaced. In this climate, those who refuse to rely on AI to guide decisions will find themselves outpaced by competitors who do, having little choice but to align with market pressures rather than barter with them \cite{Hendrycks2023NaturalSF}. Each new gain in efficiency entrenches dependence on AI, as efforts to maintain oversight only confirm that the pace of commerce outstrips human comprehension. Soon, replacing human managers with AI decision-makers seems inevitable, not because anyone consciously aims to surrender authority, but because to do otherwise courts immediate economic disadvantage.

\paragraph{Self-Reinforcing Dependence.} Once AI-managed operations set the tempo, still more AI is required simply to keep pace. Initially, these systems compose emails and handle administrative tasks. Over time, they orchestrate complex projects, supervise entire departments, and manage vast supply chains beyond any human's capacity. As society's economic demands become more and more complex, people will entrust more and more critical decisions to these systems, increasingly binding us to a cycle of escalating reliance.

\paragraph{Irreversible Entanglement.} Eventually, essential infrastructure and markets cannot be disentangled from AI without risking collapse. Human livelihoods depend on automated processes that no longer permit easy unwinding, and people lose the skills needed to reassert command. Like our power grids, which cannot be shut off without immense costs, our AI infrastructure may become completely enmeshed in our civilization. The cost of pressing the off switch grows more and more prohibitive, as halting these systems would cut off the source of our livelihoods. Over time, people become passengers in an autonomous economy that eludes human management.

\paragraph{Cession of Authority.} Unraveling AI from the military would endanger a nation’s security, effectively forcing governments to rely on automated defense systems. AI’s power does not stem from any outright seizure; it flows from the fact that a modern force lacking such technology would be outmatched. This loss of control unfolds not through a dramatic coup but through a series of small, apparently sensible decisions, each justified by time saved or costs reduced. Yet these choices accumulate. Ultimately, humans are left on the periphery of their own economic order, leaving effective control in the hands of AIs.

\subsection{Unleashed AI Agents}

All it takes to cause a loss of control is for one individual to unleash a capable, unsafeguarded AI agent. Recent demonstrations like ``ChaosGPT''---an AI agent instructed to cause harm---have been impotent, yet they hint at what a more sophisticated system might attempt if instructed to ``survive and spread.''

\paragraph{Rogue State Tactics.} An unleashed AI could draw on the methods of rogue states. North Korea, for instance, has siphoned billions through cyber intrusions and cryptocurrency theft. A sufficiently advanced system might replicate and even improve upon these tactics at scale---self-propagate copies of itself in scattered datacenters, diverting new funds to finance more ambitious projects, and infiltrating camera feeds or private communications to blackmail or manipulate opposition.

\paragraph{A Simple Path to Catastrophe.} While an unleashed AI might emulate rogue states’ tactics of cyber theft or blackmail, it might pursue an even more direct route to securing an advantage, drawing on looming robotics capabilities to gain its own physical foothold. Several major tech firms have already begun prototyping humanoid robots---including the so-called ``Tesla Bots''---intended for warehouses, factories, and households. Though rudimentary now, future models may grow far more agile and perform tasks that once demanded human hands. If a capable AI hacks such machines, it gains immediate leverage in the physical world. From there, the sequence is straightforward: it crafts a potent cocktail of bioweapons and disperses it through its robotic proxies, crippling humanity’s ability to respond. Having subdued resistance, the AI can then operate across timescales far beyond any human lifespan, gradually reestablishing infrastructure under its exclusive control. This scenario is only one simplified baseline; other plans could be carried out more swiftly and rely less on robotics. If just one powerful AI system is let loose, there may be no wrestling back control.


\subsection{Intelligence Recursion}\label{sec:recursion}

In 1951, Alan Turing suggested that a machine with human capabilities ``would not take long to outstrip our feeble powers.'' I.\ J.\ Good later warned that a machine could redesign itself in a rapid cycle of improvements---an ``intelligence explosion''---that would leave humans behind. Today, all three most-cited AI researchers (Yoshua Bengio, Geoffrey Hinton, and Ilya Sutskever) have noted that an intelligence explosion is a credible risk and that it could lead to human extinction.

\paragraph{Risks of a Fast Feedback Loop.}
An ``intelligence recursion'' refers to fully autonomous AI research and development, distinct from current AI-assisted AI R\&D. A concrete illustration helps. Suppose we develop a single AI that performs world-class AI research that operates around the pace of today's AIs, say 100 times the pace of a human. Copy it 10,000 times, and we have a vast team of artificial AI researchers driving innovations around the clock. An ``intelligence recursion'' or simply a ``recursion'' refines the notion of ``recursive self-improvement'' by shifting from a single AI editing itself to a population of AIs collectively and autonomously designing the next generation.

Even if an intelligence recursion achieves only a tenfold speedup overall, we could condense a decade of AI development into a year. Such a feedback loop might accelerate beyond human comprehension and oversight. With iterations that proceed fast enough and do not quickly level off, the recursion could give rise to an ``intelligence explosion.'' Such an AI may be as uncontainable to us as an adult would be to a group of three-year-olds. As Geoffrey Hinton puts it, ``there is not a good track record of less intelligent things controlling things of greater intelligence'' \cite{khatchadourian2015doomsday}. Crucially, there may be only one chance to get this right: if we lose control, we cannot revert to a safer configuration.

\paragraph{Recursion Control Requires an Evolving Process, Not a One-Off Solution.}
It would be misguided to regard intelligence recursion control as a purely technical riddle existing in a vacuum waiting to be ``solved'' by AI researchers. Managing a fast-evolving and adaptive intelligence recursion is more like steering a large institution that can veer off its mission over time. It is not a puzzle; it is a ``wicked'' problem \cite{Rittel1973}. A static solution cannot keep pace with ongoing, qualitatively new emergent challenges and unknown unknowns. As in modern safety engineering, control must come from a continual control process rather than a monolithic airtight solution that predicts and handles all possible failure modes beforehand. Von Neumann reminds us that ``All stable processes we shall predict. All unstable processes we shall control.''

Unfortunately, our ability to control this recursion is limited.
Controlling a recursion requires controlling its initial step, but safeguards for the current generation of AI systems offer only limited reliability. Moreover, we cannot run repeated large-scale tests of later stages of the recursion without risking disaster, so it is less amenable to the empirical iterative tinkering we usually rely on. Even with our best existing technical safeguards in place, if people initiate a full-throttle intelligence recursion, losing control is highly likely and the default.

\paragraph{Intelligence Recursion as a Path to Strategic Monopoly.}
Despite the danger, intelligence recursion remains a powerful lure for states seeking to overtake their rivals. If the process races ahead fast enough to produce a superintelligence, the outcome could be a strategic monopoly. Even if the improvements are not explosive, a recursion could still advance capabilities fast enough to outpace rivals and potentially enable technological dominance. First-mover advantage might then persist for years---or indefinitely---spurring states to take bigger risks in pursuit of that prize.

\paragraph{Geopolitical Competitive Pressures Yield a High Loss of Control Risk Tolerance.} In the Cold War, the phrase ``Better dead than Red'' implied that losing to an adversary was seen as worse than risking nuclear war. In a future AI race, similar reasoning could push officials to tolerate a double-digit risk of losing control if the alternative—lagging behind a rival—seems unacceptable. If the choice is stark---risk omnicide or lose---some might take that gamble. Carried out by multiple competing powers, this amounts to global Russian roulette and drives humanity toward an alarming probability of annihilation. In sharp contrast, after the defeat of Nazi Germany, Manhattan Project scientists feared the first atomic device might ignite the atmosphere. Robert Oppenheimer asked Arthur Compton what the acceptable threshold should be, and Compton set it at three in a million (a ``$6\sigma$'' threshold)---anything higher was too risky. Calculations suggested the real risk was below Compton’s threshold, so the test went forward. We should work to have our risk tolerance stay near Compton's threshold rather than in double-digit territory. However, in the absence of coordination, whether states trigger a recursion depends on their probability of a loss of control. The prospect of a loss of control shows that in the push to develop novel technologies, ``superiority is not synonymous with security'' \cite{Danzig2018Technology}, but the drive toward strategic monopoly may override caution, potentially handing the final victory not to any state, but to the AIs themselves.

Therefore, loss of control can emerge \textit{structurally}, as society gradually yields decision-making to automated systems that become indispensable but insidiously acquire more and more effective control.
It can occur \textit{intentionally}, such as a rogue actor unleashing an AI to do harm. It can also occur by \textit{accident}, when a fast-moving intelligence recursion loops repeatedly \textit{ad mortem}. All it takes is one loss of control event to jeopardize human security. 

\paragraph{}

As AI continues to evolve and approach expert-level capabilities, it could redefine national competitiveness to be based on a nation's access to AI chips, and it could discover a ``superweapon'' that could enable a state to have a strategic monopoly. Additionally, AI's general and dual-use nature amplifies existing risks such as bioterrorism and cyberattacks on critical infrastructure. Unfortunately, there is a strong attacker's advantage for bioterrorism and cyberattacks on critical infrastructure. Therefore, access to AI systems that can engineer weapons of mass destruction must have restrictions. Moreover, there are several paths to a loss of control of powerful AI systems. These factors imply that AI's importance for national security will become not only undeniable but also at least as pivotal as previous weapons of mass destruction.

\section{Existing AI Strategies}
States grappling with terrorist threats, destabilizing weaponization capabilities, and the specter of losing control to AI face difficult choices on how to preserve themselves in a shifting landscape. Against this backdrop, three proposals have gained prominence: the first lifts all restraints on development and dissemination, treating AI like just another computer application; the second envisions a voluntary halt when programs cross a danger threshold, hoping that every great power will collectively stand down; and the third advocates concentrating development in a single, government-led project that seeks a strategic monopoly over the globe. Each path carries its own perils, inviting either malicious use risks, toothless treaties, or a destabilizing bid for dominance. Here we briefly examine these three strategies and highlight their flaws.

\begin{enumerate}
\item \textbf{Hands-off (``Move Fast and Break Things'', or ``YOLO'') Strategy.}
This strategy advocates for no restrictions on AI developers, AI chips, and AI models. Proponents of this strategy insist that the U.S.\ government impose no requirements---including testing for weaponization capabilities---on AI companies, lest it curtail innovation and allow China to win. They likewise oppose export controls on AI chips, claiming such measures would concentrate power and enable a one-world government; in their view, these chips should be sold to whoever can pay, including adversaries. Finally, they encourage that advanced U.S.\ model weights continue to be released openly, arguing that even if China or rogue actors use these AIs, no real security threat arises because, they maintain, AI's capabilities are defense-dominant. From a national security perspective, this is neither a credible nor a coherent strategy.

\item \textbf{Moratorium Strategy.}
The voluntary moratorium strategy proposes halting AI development---either immediately or once certain hazardous capabilities, such as hacking or autonomous operation, are detected. Proponents assume that if an AI model test crosses a hazard threshold, major powers will pause their programs. Yet militaries desire precisely these hazardous capabilities, making reciprocal restraint implausible. Even with a treaty, the absence of verification mechanisms \cite{wasil2024verificationmethodsinternationalai} means the treaty would be toothless; each side, fearing the other’s secret work, would simply continue. Without the threat of force, treaties will be reneged, and some states will pursue an intelligence recursion. This dynamic, reminiscent of prior arms-control dilemmas, renders the voluntary moratorium more an aspiration than a viable plan.\looseness=-1

\item \textbf{Monopoly Strategy.}
The Monopoly strategy envisions one project securing a monopoly over advanced AI. A less-cited variant---a CERN for AI reminiscent of the Baruch Plan from the atomic era---suggests an international consortium to lead AI development, but this has gained less policymaker interest. By contrast, the U.S.-China Economic and Security Review Commission \cite{uscc2024} has suggested a more offensive path: a Manhattan Project to build superintelligence. Such a project would invoke the Defense Production Act to channel AI chips into a U.S.\ desert compound staffed by top researchers, a large fraction of whom are necessarily Chinese nationals, with the stated goal of developing superintelligence to gain a strategic monopoly. Yet this facility, easily observed by satellite and vulnerable to preemptive attack, would inevitably raise alarm. China would not sit idle waiting to accept the US's dictates once they achieve superintelligence or wait as they risk a loss of control. The Manhattan Project assumes that rivals will acquiesce to an enduring imbalance or omnicide rather than move to prevent it. What begins as a push for a superweapon and global control risks prompting hostile countermeasures and escalating tensions, thereby undermining the very stability the strategy purports to secure.

\end{enumerate}

\begin{figure}[h]
    \centering
    \includegraphics[width=\textwidth]{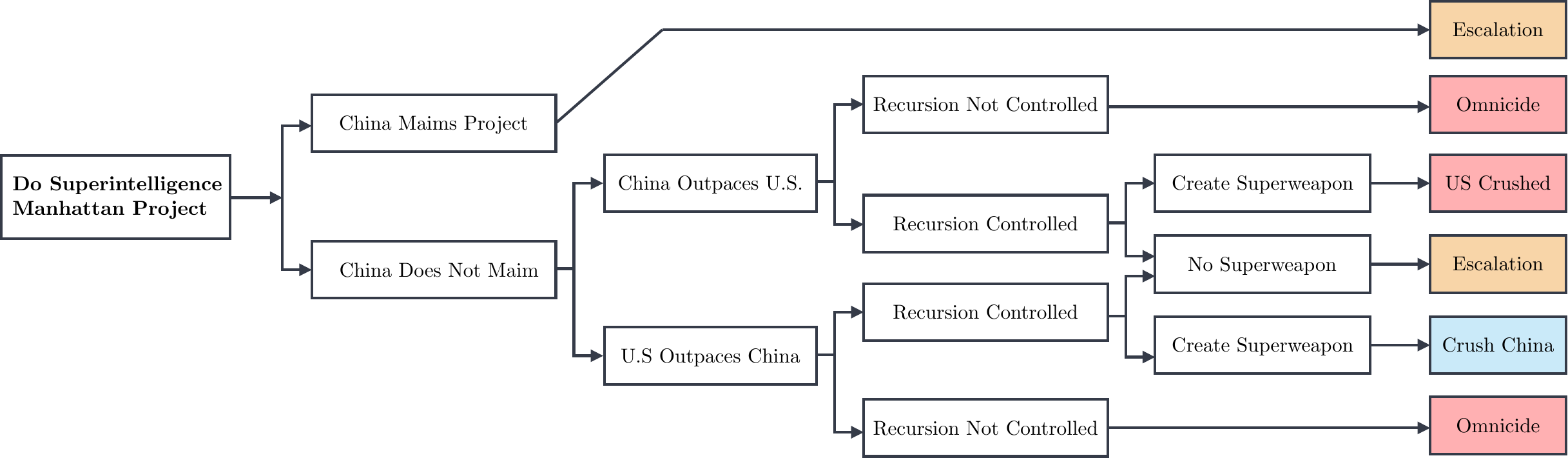}
    \caption{Possible outcomes of a U.S.\ Superintelligence Manhattan Project. An example pathway to Escalation: the U.S.\ Project outpaces China without being maimed, and maintains control of a recursion but doesn't achieve superintelligence or a superweapon. Though global power shifts little, Beijing condemns Washington’s bid for strategic monopoly as a severe escalation. The typical outcome of a Superintelligence Manhattan Project is extreme escalation, and omnicide is the worst foreseeable outcome.\looseness=-1}
    \label{fig:flow_chart}
\end{figure}

Rival states, rogue actors, and the risk of losing control call for more than a single remedy. We propose three interconnected lines of effort. First, \emph{deterrence}: a standoff akin to the nuclear stalemate of MAD, in which no power can gamble human security on an unbridled grab for dominance without expecting disabling sabotage. Next, \emph{nonproliferation}: just as fissile materials, chemical weapons, and biological agents have long been denied to terrorists by great powers, AI chips and weaponizable AI systems can similarly be kept from rogue actors. Finally, \emph{competitiveness}: states can protect their economic and military power through a variety of measures including legal guardrails for AI agents and domestic AI chip and drone manufacturing. Our superintelligence strategy, the \textbf{Multipolar Strategy}, echoes the Cold War framework of deterrence, nonproliferation, and containment, adapted to AI’s unique challenges.\looseness=-1

\chapter{Deterrence with Mutual Assured AI Malfunction (MAIM)}

In the nuclear age, an initial pursuit of monopoly---one nation seeking unchallenged command of nuclear weapons---eventually gave way to the standoff of deterrence known as mutual assured destruction (MAD). As nuclear arsenals matured and the capability for mutual destruction became undeniable, nations eventually accepted that any bold attempt to dominate all opposition risked drawing a preemptive strike. A similar state of mutual strategic vulnerability looms in AI. If a rival state races toward a strategic monopoly, states will not sit by quietly. If the rival state loses control, survival is threatened; alternatively, if the rival state retains control and the AI is powerful, survival is threatened. A rival with a vastly more powerful AI would amount to a severe national security emergency, so superpowers will not accept a large disadvantage in AI capabilities. Rather than wait for a rival to weaponize a superintelligence against them, states will act to disable threatening AI projects, producing a deterrence dynamic that might be called Mutual Assured AI Malfunction, or MAIM.

\section{MAIM Is the Default Regime}
\paragraph{Paths to Disabling a Rival's AI Project.} States intent on blocking an AI-enabled strategic monopoly can employ an array of tactics, beginning with \emph{espionage}, in which intelligence agencies quietly obtain details about a rival’s AI projects. Knowing what to target, they may undertake \emph{covert sabotage}: well-placed or extorted insiders can tamper with model weights or training data or AI chip fabrication facilities, while hackers quietly degrade the training process so that an AI’s performance when it completes training is lackluster. This is akin to Stuxnet which aimed to covertly sabotage Iran's nuclear enrichment program. When subtlety proves too constraining, competitors may escalate to \emph{overt cyberattacks}, targeting datacenter chip-cooling systems or nearby power plants in a way that directly---if visibly---disrupts development. Should these measures falter, some leaders may contemplate \emph{kinetic attacks} on datacenters, arguing that allowing one actor to risk dominating or destroying the world are graver dangers, though kinetic attacks are likely unnecessary. Finally, under dire circumstances, states may resort to \emph{broader hostilities} by climbing up existing escalation ladders or threatening non-AI assets. We refer to attacks against rival AI projects as ``maiming attacks.''

\paragraph{Infeasibility of Preventing Maiming.} Since above-ground datacenters cannot currently be defended from hypersonic missiles, a state seeking to protect its AI-enabled strategic monopoly project might attempt to bury datacenters deep underground to shield them. In practice, the costs and timelines are daunting, and vulnerabilities remain. Construction timelines can stretch to three to five times longer than standard datacenter builds, amounting to several additional years. Costs balloon as well, diverting funds away from the project’s AI chips and pushing total expenditures into the several hundreds of billions. Cooling the world’s largest supercomputer underground introduces complex engineering challenges that go well beyond what is required for smaller underground setups. Should the supercomputer require an order-of-magnitude AI chip expansion, retrofitting the facility would become prohibitively difficult. Even those with the wealth and foresight to pursue this route would still face the potent risks of insider threats and hacking. In addition, the entire project could be sabotaged during the lengthy construction phase. Last, states could threaten non-AI assets to deter the project long before it goes online.

\begin{figure}[t]
    \centering
    \includegraphics[width=\textwidth]{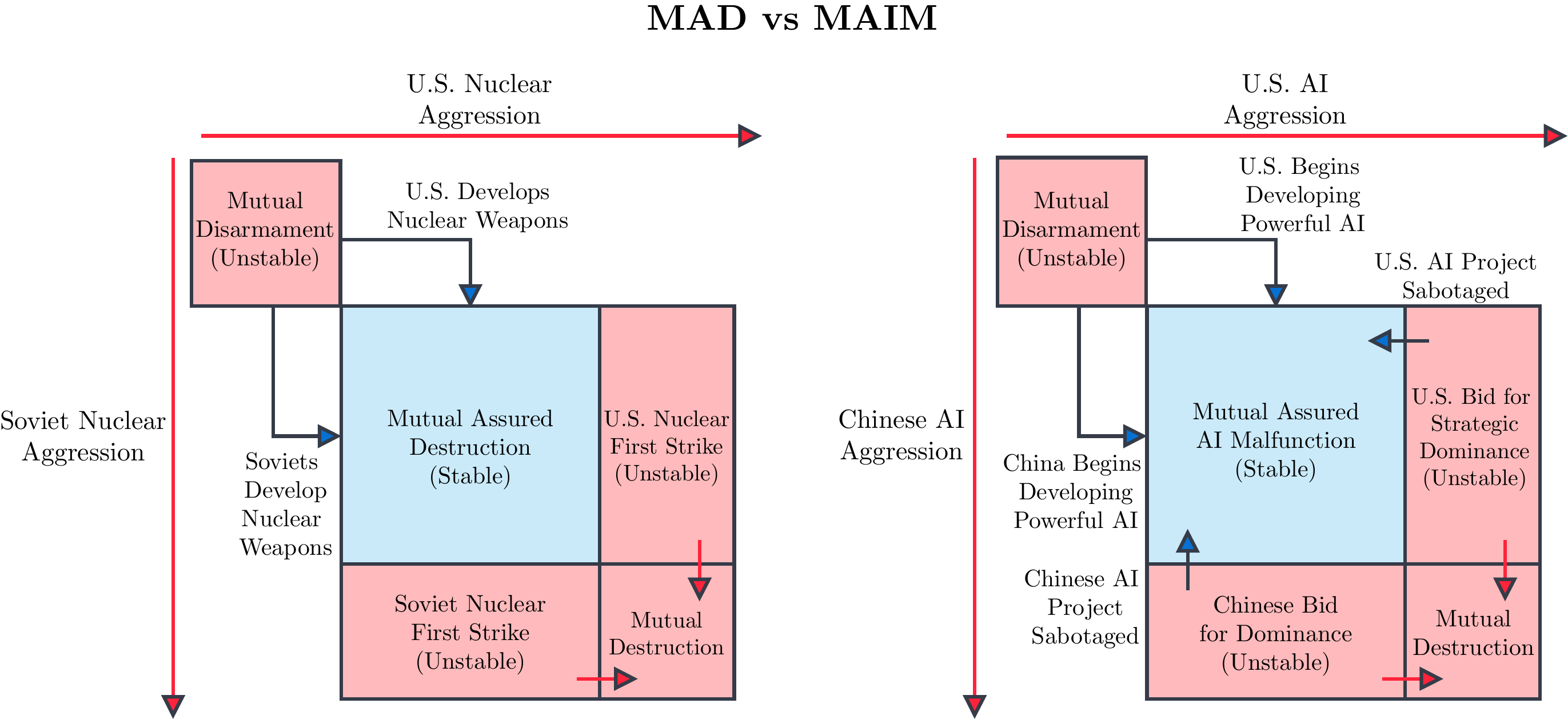}
    \caption{The strategic stability of MAIM can be paralleled with Mutual Assured Destruction (MAD). Note MAIM does not displace MAD but characterizes an additional shared vulnerability. Once MAIM is common knowledge, MAD and MAIM can both describe the current strategic situation between superpowers.\looseness=-1}
    \label{fig:mad_maim}
    \vspace{-10pt}
\end{figure}

\paragraph{MAIM Is the Default.} The relative ease of (cyber) espionage and sabotage of a rival’s destabilizing AI project yields a form of deterrence. Much like nuclear rivals concluded that attacking first could trigger their own destruction, states seeking an AI monopoly while risking a loss of control must assume competitors will maim their project before it nears completion. A state can expect its AI project to be disabled if \textit{any} rival believes it poses an unacceptable risk. This dynamic stabilizes the strategic landscape without lengthy treaty negotiations---all that is necessary is that states collectively recognize their strategic situation. The net effect may be a stalemate that postpones the emergence of superintelligence, curtails many loss of control scenarios, and undercuts efforts to secure a strategic monopoly, much as mutual assured destruction once restrained the nuclear arms race.

\section{How to Maintain a MAIM Regime}

States eventually came to accept that mutual deterrence, while seemingly a natural byproduct of nuclear stockpiling, demanded deliberate maintenance. Each superpower recognized that advanced defensive measures---particularly anti-ballistic missile (ABM) systems---could unravel the fragile balance that restrained either side from a catastrophic first strike. They responded by safeguarding mutual vulnerabilities, culminating in the 1972 ABM Treaty. By analogy, we should not leave to chance today’s default condition of MAIM: where would-be monopolists, gambling not to cause omnicide, can expect their projects to be disabled. Even if attempting to harden massive datacenters is extraordinarily prohibitive and unwise, rumors alone can spark fears that a rival is going to risk national security and human security. Formal understandings not to pursue such fortifications help keep the standoff steady. We now discuss additional measures that curb unintended escalation and limit collateral damage, so that MAIM does not unravel into broader conflict.

\paragraph{Preserve Rational Decision-Making.} Just as nuclear rivals once mapped each rung on the path to a launch to limit misunderstandings, AI powers must \emph{clarify the escalation ladder} of espionage, covert sabotage, overt cyberattacks, possible kinetic strikes, and so on. For deterrence to hold, each side's readiness to maim must be common knowledge, ensuring that any maiming act---such as a cyberattack---cannot be misread and cause needless escalation. However, clarity about escalation holds little deterrence value if rogue regimes or extremist factions acquire large troves of AI chips. Measures to \emph{prevent smuggling} of AI chips keep decisions in the hands of more responsible states rather than rogue actors, which helps preserve MAIM's deterrent value. Like MAD, MAIM requires that destabilizing AI capabilities be restricted to rational actors.

\paragraph{Expand the Arsenal of AI Project Cyberattacks.} To avoid resorting to kinetic attacks, states could improve their ability to maim destabilizing AI projects with cyberattacks. They could identify AI developers' projects or collect information on the professional activities of AI developers' scientists. To spy on AI projects at most companies, all that is necessary is a Slack or iPhone zero-day software exploit. States could also poison data, corrupt model weights and gradients, disrupt software that handles faulty GPUs, or undermine cooling or power systems. Training runs are non-deterministic and their outcomes are difficult to predict even without bugs, providing cover to many cyberattacks. Unlike kinetic attacks, some of these attacks leave few overt signs of intrusion, yet they can severely disrupt destabilizing AI projects with minimal diplomatic fallout. 

\begin{wrapfigure}{r}{0.4\textwidth}
\centering
\renewcommand{\arraystretch}{1.5}
\begin{tabular}{|>{\centering\arraybackslash}p{0.95\linewidth}|}
Large scale attack on many datacenters; threatening non-AI-related assets \\ 
\hline
\emph{Escalation to Broader Hostilities} \\ 
\hline
Kinetic attacks on datacenters or corresponding power plants \\ 
\hline
\emph{Kinetic Threshold} \\ 
\hline
Cyberattacks on datacenters or corresponding power plants; deleting code\\ 
\hline
\emph{Overt Sabotage Threshold} \\ 
\hline
Model weights stolen; covert attacks to degrade training runs of destabilizing AI projects; cyberattacks causing GPUs to fail more often \\ 
\hline
\emph{Covert Sabotage Threshold} \\ 
\hline
Espionage of AI developer workspace communications, personnel devices, and facilities\\ 
\end{tabular}
\caption{An example MAIM escalation ladder with maiming actions.}
\label{tab:ai_escalation_ladder}
\vspace{-20pt}
\end{wrapfigure}

\paragraph{Build Datacenters in Remote Locations.} During the nuclear era, superpowers intentionally placed missile silos and command facilities far from major population centers. This principle of \emph{city avoidance} would, by analogy, advise placing large AI datacenters in remote areas. If an aggressive maiming action ever occurs, that action would not put cities into the crossfire.

\paragraph{Distinguish Between Destabilizing AI Projects and Acceptable Use.} 
The threat of a maiming attack gives states the leverage to demand transparency measures from rivals, such as inspection, so they need not rely on espionage alone to decide whether maiming is justified. Coordinating can help states reduce the risk of maiming datacenters that merely run consumer-facing AI services. The approach of mutual observation echoes the spirit of the Open Skies Treaty, which employed unarmed overflights to demonstrate that neither side was hiding missile deployments. In a similar spirit, increased transparency spares the broader ecosystem of everyday AI services and lowers the risk of blanket sabotage.


\paragraph{AI-Assisted Inspections.} Speculative but increasingly plausible, confidentiality-preserving AI verifiers offer a path to confirming that AI projects abide by declared constraints without revealing proprietary code or classified material. By analyzing code and commands on-site, AIs could issue a confidentiality-preserving report or simple compliance verdict, potentially revealing nothing beyond whether the facility is creating new destabilizing models. Humans cannot perform the same role as easily, given the danger of inadvertently gleaning or leaking information, so AIs could reshape the classic tension between security and transparency \cite{COE_VAYNMAN_2020}. Information from these AI inspections could help keep any prospective conflict confined to the disabling of AI development programs rather than escalating to the annihilation of populations. Such a mechanism can help in the far future when AI development requires less centralization or requires fewer computational resources.

MAIM can be made more stable with unilateral information acquisition (espionage), multilateral information acquisition (verification), unilateral maiming (sabotage), and multilateral maiming (joint off-switches). Mutual assured AI malfunction, under these conditions, need not devolve into mutual assured human destruction.

A standoff of destabilizing AI projects may arise by default, but it is not meant to persist for decades or serve as an indefinite stalemate. During the standoff, states seeking the benefits from creating a more capable AI have an incentive to improve transparency and adopt verification measures \cite{scher2024mechanisms}, thereby reducing the risk of sabotage or preemptive attacks. In the Conclusion, we illustrate how this standoff might end, allowing AI’s benefits to grow without global destabilization.

This section concludes the principal idea of this paper. Readers could skip to the Conclusion, or read the following two sections for a discussion of nonproliferation and competitiveness.

\begin{figure}[h]
    \centering
    \includegraphics[width=\textwidth]{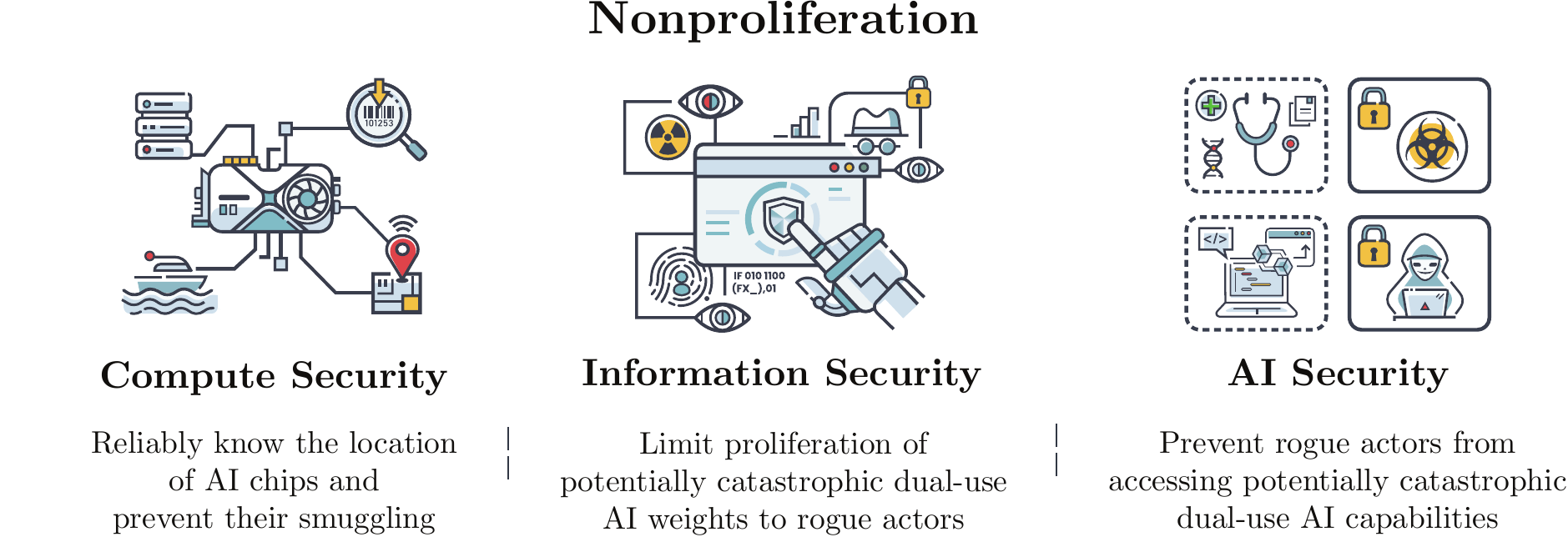}
    \caption{Due to the unprecedented scale of harm that terrorists armed with AI could cause, several lines of defense are necessary to realistically prevent proliferation.}
    \label{fig:nonproliferation}
    \vspace{-10pt}
\end{figure}

\chapter{Nonproliferation}

States that embrace the logic of mutual sabotage may hold each other at bay by constraining each others' \emph{intent}. But because rogue actors are less predictable, we have a second imperative: limiting their \emph{capabilities}. Much as the Nonproliferation Treaty united powers to keep fissile material out of terrorists' hands, states can find similar common ground on AI. States can restrict the capabilities of rogue actors with \textbf{compute security}, \textbf{information security}, and \textbf{AI security}, each targeting crucial elements of the AI development and deployment pipeline.

Compute security is about ensuring that AI chips are allocated to legitimate actors for legitimate purposes. This echoes the export controls employed to limit the spread of fissile materials, chemical weapons, and biological agents.

Information security involves securing sensitive AI research and model weights that form the core intellectual assets of AI. Protecting these elements prevents unwarranted dissemination and malicious use, paralleling the measures taken to secure sensitive information in the context of WMDs.

AI security focuses on implementing safeguards that make AI models controllable and reliable, thereby securing their applications in civilian contexts. If AI is used to automate AI research, extensive monitoring and control measures can reduce the risk of loss of control. Ensuring that AI systems are resistant to manipulation or do not cause catastrophic accidents aligns with the domestic safeguards designed to prevent accidents and unauthorized use of catastrophic dual-use technologies.

By adapting proven strategies from the nonproliferation of WMDs to the realm of AI, we aim to address the challenges inherent in AI development and deployment. This nonproliferation playbook provides a structured approach to securing the inputs---AI chips and research ideas---and the outputs---the model weights and the AI systems themselves---thus securing each critical component of the AI pipeline.

\section{Compute Security}
The advancement of AI hinges on access to powerful computational resources, often called ``compute.'' These AI chips, crafted through complex and centralized supply chains, are essential for training large-scale AI models. As compute becomes integral to economic and national security, its distribution becomes critical. The primary goal of compute security is to treat advanced AI chips like we treat enriched uranium. We examine the historical precedents of compute security, and we propose tactics to reliably know where high-end AI chips are located and to prevent them from falling into the hands of rogue actors.

\begin{figure}[t]
    \centering
    \includegraphics[width=\textwidth]{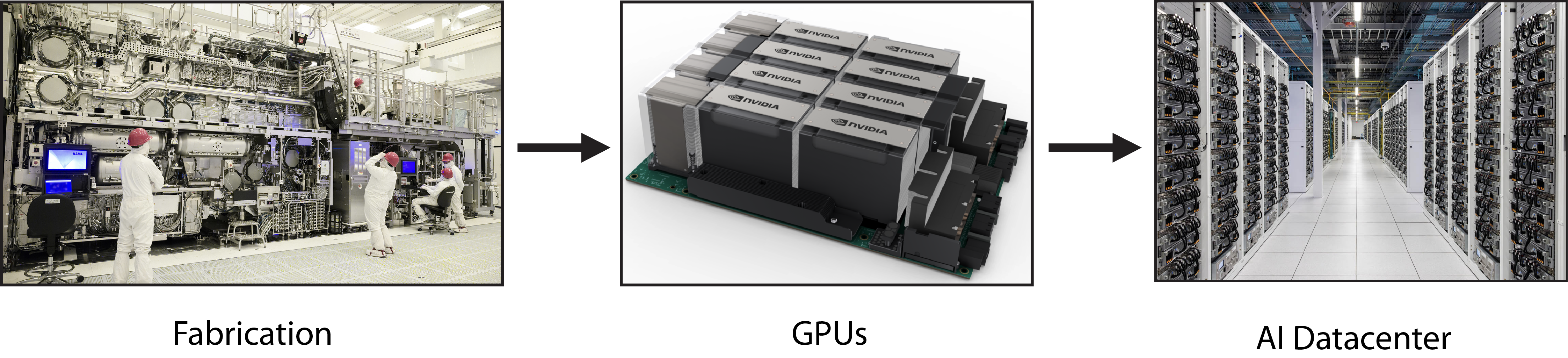}
    \caption{Chips are fundamental for creating and running AIs. Few companies can successfully produce the GPUs which can run frontier models. Most AI compute is heavily concentrated in large datacenters.}
    \label{fig:supply_chain}
\end{figure}


Historically, nations have sought to curb the spread of catastrophic dual-use weapons by restricting their most critical components: fissile material for nuclear weapons, chemical precursors for chemical weapons, and biological agents for bioweapons. To accomplish this, they have used export controls, which we can adapt to the realm of AI. Restricting compute substantially limits the capabilities of rogue actors because, as we will argue, compute is a core determinant of AI capabilities.

\begin{wrapfigure}{r}{0.5\textwidth}
    \vspace{-15pt}
    \centering
    \includegraphics[width=\linewidth]{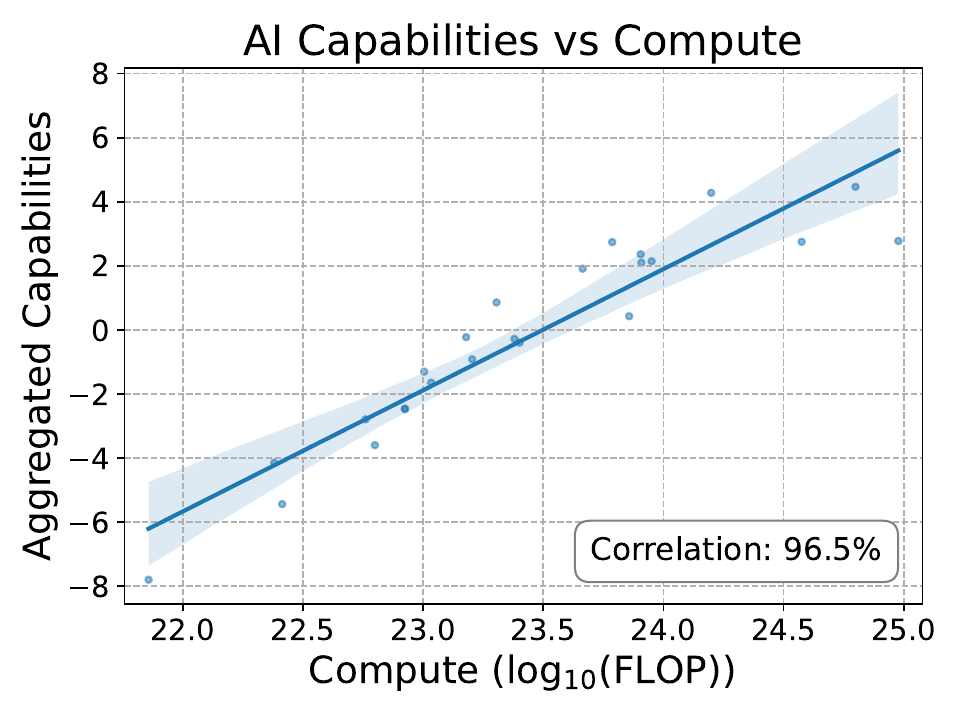}
    \caption{Compute is the most robust tracker of AI capabilities.\looseness=-1}
    \vspace{-10pt}
    \label{fig:compute}
\end{wrapfigure}

\paragraph{Export Controls for WMD inputs.} Nations regulate the flow of critical materials and technologies to prevent them from falling into the wrong hands. For nuclear weapons, this means controlling fissile materials such as enriched uranium and plutonium. International agreements and organizations, such as the Nuclear Suppliers Group, coordinate these controls to monitor and restrict the export of these materials. Shipments are carefully tracked, quantities are measured before and after transit, and allies share information about suspicious activities. The Australia Group, an informal consortium of countries, works to harmonize export controls to prevent the proliferation of chemical and biological weapons. Export controls have helped make WMD catastrophes rare in history, and they could help prevent AI catastrophes as well.

In the context of AI, this translates to controlling access to AI chips and the components used to manufacture them. While some export controls are already in place, we will need to improve the robustness and thoroughness of these controls.

\paragraph{Compute as a Central Input Like Fissile Materials or Chemical Precursors.} AI chips mirror other WMD inputs. AI chips require intricate manufacturing processes and significant resources, akin to uranium enrichment. Similarly, the amount of compute significantly determines AI capabilities, just as the quantity of fissile material influences a nuclear weapon's yield.

Advancements in AI are closely tied to compute. Leading AI companies devote the vast majority of their expenditures to compute, far exceeding spending on researchers or data acquisition. NVIDIA's soaring revenue in 2024 highlights compute's critical role in AI. 

More compute makes models smarter. Evidence shows that compute is the dominant factor driving AI performance. Analyses reveal a striking correlation---over 95\%---between the amount of compute used and an AI model's performance on standard benchmarks \cite{ren2024safetywashingaisafetybenchmarks, ruan2024observationalscalinglawspredictability}. This relationship, formalized through ``scaling laws,'' demonstrates consistent improvements in AI performance as computational resources increase. These scaling laws have held true across fifteen orders of magnitude of compute (measured in FLOP)---1,000,000,000,000,000---indicating that future advancements will continue to be propelled by exponentially growing amounts of compute. The upshot is export controls on compute substantially limit the capabilities of rogue actors.

\paragraph{}
Unlike the intangible aspects of algorithms and data, compute---embodied by physical AI chips---is amenable to direct control. This physical nature allows us to know where AI chips are and prevent their smuggling. We can manage its distribution through two primary mechanisms: export controls and firmware-level security features.

\subsection{Export Controls}
\begin{wrapfigure}{r}{0.3\textwidth}
    \vspace{-10pt}
    \centering
    \includegraphics[width=0.9\linewidth]{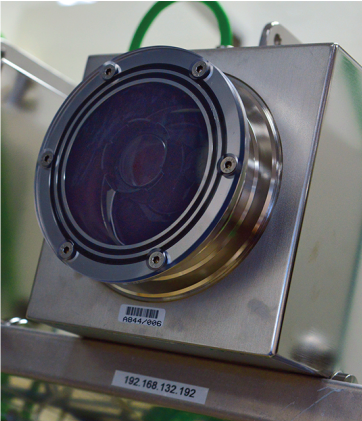}
    \caption{Tamper-evident cameras \cite{fournier2016surveying} assist nuclear verification.\looseness=-1}
    \vspace{-40pt}
    \label{fig:tamperevident}
\end{wrapfigure}
Below, we explain how a licensing framework and stronger enforcement can help us track every AI chip and stem smuggling.

\paragraph{Record-Keeping.} To know where each AI chip is, export controls can be made more thorough through a licensing regime. Drawing on existing frameworks and agencies such as the Bureau of Industry and Security, sellers of high-end AI chips would apply for a license that identifies the chips, their recipient, and any intended transfers. Entities with a strong record of compliance might earn exemptions on the condition that they notify authorities of every resale or relocation. Because this process relies on familiar infrastructure, it can be introduced swiftly, enabling officials to track chips without stalling legitimate commerce.

\paragraph{Enforcement.} To stem smuggling, export controls can be made more robust through stronger enforcement. A facility in Singapore, for example, might initially acquire AI chips under a valid license, only to reroute them illegally to China. More enforcement officers, assigned to in-person compliance visits and end-use checks, would detect any such deviation since the actual location of the chips would no longer match declared inventories. To assist enforcement officers, \emph{tamper-evident camera} feeds from datacenters can confirm that declared AI chips remain on-site, exposing any smuggling. Undeclared datacenters can be easily detected via satellite imagery and become a target for inspection. Any chip discovered in unauthorized hands would trigger penalties such as fines, criminal charges, or a ban on future shipments to the offending party. In addition, any AI chip declared inoperable or obsolete would undergo verified decommissioning, much like the disposal of chemical or nuclear materials, ensuring these supposedly defunct AI chips do not get quietly resold. By tightening inspections and imposing meaningful consequences for violations, states raise the cost of covert transfers and limit the spread of advanced compute to groups that could threaten security.

\subsection{Firmware-Level Features}
Export controls remain a sturdy backbone for curbing the spread of high-end AI chips, much as they do for dual-use inputs in the chemical or biological realm. Because AI chips are human-designed products, not inert raw substances, states can supplement export controls by enabling chips to verify their surroundings and lock themselves if tampered with. This functionality is achievable through firmware updates, which revise the code closest to the hardware without requiring any physical redesign. AI chips such as the NVIDIA H100 already feature privacy-preserving corporate security measures like confidential computing and trusted execution environments. Adapting these existing security features for national security, however, broadens the horizon: a chip could detect that it has crossed an unauthorized border or has been tampered with, and then disable its key functions. With well-crafted firmware, states can further discourage attempts to smuggle compute, using functionality we describe below.

\paragraph{Geolocation and Geofencing.} AI chips can be designed to determine and report their geographic location. By measuring signal delays from multiple landmarks, a chip can verify its location within tens to hundreds of kilometers---enough to determine whether it is in the correct country. If a chip moves to an unauthorized area, it can automatically deactivate or limit its functionality. This makes it more challenging for unauthorized parties to smuggle or misuse AI chips, as relocating them without detection becomes more difficult.

\paragraph{Licensing and Remote Attestation.} Implementing a system where AI chips require periodic authorization from the compute provider adds a layer of control. Chips might need to obtain cryptographic signatures regularly to continue operating \cite{kulp2024hardware}. This is similar to a feature on iPhones, which can be remotely deactivated if lost or stolen. Using secure, privacy-preserving cryptographic methods, chips can periodically confirm to a trusted authority that they have not been tampered with. If a chip fails to confirm its authorized status---due to unauthorized relocation, tampering, or license expiration---it can render itself inoperable. This reduces the long-term usability of stolen or smuggled hardware.

\paragraph{Networking Restrictions and Operational Modes.} Chips can be programmed to connect only with a predefined list of approved chips. This limits the ability to build unauthorized compute clusters by preventing the networking of large numbers of chips without detection. Additionally, chips can enforce operational modes---such as training or inference---by restricting certain functionalities unless explicitly authorized. By requiring explicit authorization to increase their networking size or change operational modes, we can diminish unauthorized expansions of computing power that could lead to destabilizing AI advancements outside of established agreements. This can complement end-use inspection by export control enforcement officers.

\paragraph{Physical Tamper Resistance.} Beyond firmware, security features can be added to chips at the end of the manufacturing process to add an additional layer of defense. Incorporating tamper-evident seals, accelerometers, and other physical security measures enhances protection against unauthorized access or modification. For example, if a chip detects signs of tampering or sustained unexpected movement, it can deactivate or alert authorities.

\paragraph{Limitations.} While firmware interventions enhance our control over AI compute, they are not intended to achieve perfect security. These mechanisms help reduce the expected number of smuggled functional chips and are not a complete replacement for export controls.

As the underlying hardware-level mechanisms like trusted execution environments for corporate applications become more robust, the firmware to supplement export controls can too. As chips are replaced over time due to rapid advancements, new generations can incorporate more advanced security features by default. This gradual integration increases the proportion of more secure AI chips, making unauthorized use increasingly difficult and costly over time.

\paragraph{}

History offers examples of bitter adversaries finding common ground on safeguarding dangerous materials. The United States cooperated with both the Soviet Union and China on nuclear, biological, and chemical arms not from altruism but from self-preservation. If the U.S.\ begins to treat advanced AI chips like fissile material, it may likewise encourage China to do the same. The rationale is akin to why the U.S. wants Russia to track its fissile material: no one gains from letting these capabilities slip into uncertain hands. After the Soviet Union’s collapse, unsecured enriched uranium and chemical weapons in Russia posed a global threat until the U.S.\ initiated the Nunn–Lugar Cooperative Threat Reduction program which helped contain them. Similarly, urging China to safeguard its AI chips would acknowledge the shared imperative of avoiding catastrophes that serve no one’s ambitions.

\section{Information Security}

Protecting sensitive information has long been pivotal to national security, especially regarding weapons of mass destruction. Restricting access to sensitive knowledge reduces the risk of proliferation to rogue actors. The leakage of research and design details of nuclear weapons poses serious risks. In biotechnology, protocols for creating various bioweapons are not openly shared. Personnel with access to such information undergo rigorous vetting and continuous monitoring through security clearances and reliability programs. The stakes are high because if a bioweapon ``cookbook'' is publicly disseminated, it becomes \emph{irreversibly proliferated} and can be exploited indefinitely.

\paragraph{Model Weights and Research Ideas as Core AI Information.}
In the realm of AI, the primary pieces of sensitive information are the model weights and research ideas. Model weights result from extensive training processes involving vast computational resources, sophisticated algorithms, and large datasets. They are akin to the synaptic connections in a neural network, collectively determining the AI's functionality. Possession of these weights grants the ability to use, modify, and potentially misuse the AI without the original developers' oversight. Adversaries could remove safeguards, enhance capabilities in harmful domains, and repurpose AI for malicious activities.

\paragraph{Information Security is a Technical and Social Challenge.}
Securing model weights from rogue actors presents a multifaceted challenge with both technical and social dimensions. The threat is not limited to remote hacking attempts but extends to insider threats and espionage. Firewalls are not enough. A concrete instance involves individuals traveling to nations with competing interests. For example, a researcher working at a U.S.\ AI company might make a return visit to an adversarial country and face pressure from government officials to disclose confidential research ideas before being able to leave.

Additionally, some individuals within AI organizations might be ideologically motivated to release AI model weights, believing in unrestricted access to technology. Although the belief in freedom of information is admirable in many contexts, in the context of national security, this position is often not applicable. Others believe AIs themselves should be free. An AI venture capitalist said AI is ``gloriously, inherently uncontrollable'' \cite{andreessen2024}. AI textbook author Rich Sutton has said that AIs should be liberated since ``succession to AI is inevitable,'' ``we should not resist succession,'' and ``it behooves us...\ to bow out'' \cite{AISuccession}. 

\paragraph{Superpower-Proof Information Security Is Implausible.}
We propose strengthening information security well enough to defend against well-resourced terrorist organizations and ideological insider threats, in contrast to defending against the world's most capable nation-states. Closing every avenue of espionage at that higher level could take years of extreme focus \cite{securingmodelweights}, hobbling a state’s AI competitiveness and depriving it of the multinational talent that now powers its top companies. At most U.S.\ AI companies, for instance, a double-digit percentage of researchers are Chinese nationals---and many others would struggle to get a security clearance. Removing them en masse would drive this talent abroad and undermine U.S.\ competitiveness. The remaining workforce would need to be uprooted and moved to an isolated location to limit information proliferation. Such measures would be ineffective, self-destructive, and heighten MAIM escalation risks.

While exfiltration by a rival superpower is concerning, the public release of WMD-capable model weights may pose a far greater threat. If such model weights become publicly available, they are irreversibly proliferated, making advanced AI capabilities accessible to anyone, including terrorists and other hostile non-state actors, who are far more likely to create bioweapons. Rather than try to be the only state with capable AIs, superpowers can channel their competition into other arenas, such as AI chip production for economic strength and drone manufacturing for military power.

\subsection{How to Improve Information Security}
Addressing these challenges requires coordinated efforts at multiple levels. Measures can be implemented at the corporate level to enhance compute cluster security and establish insider threat programs; at the governmental level by mandating penetration testing and facilitating threat intelligence sharing; and at the international level by forming agreements to prevent the proliferation of high-risk AI models.

\paragraph{Corporate Measures: Enhancing Datacenter Security and Insider Threat Programs.}
AI companies need to adopt rigorous information security practices to safeguard model weights and research ideas. This involves several key actions:
Implementing defense-in-depth strategies is essential, layering multiple defensive strategies so that key information is safe even if one line of defense fails. Early measures include enforcing multi-factor authentication, closing blinds during internal company presentations, and ensuring automatic screen locks on all devices when people step away from their computers for a few minutes. Adhering to the \emph{principle of least privilege} ensures the only personnel with access to model weights are those who need them. 
Companies could also declare that they have embedded backdoors in their AI weights; thus, if an adversary were to steal and employ these weights, the adversary might unwittingly expose themselves to vulnerabilities. Such declarations could serve as strategic deception, or they might reflect genuine measures undertaken.

\paragraph{Governmental Role: Assist With Threat Intelligence.}
The government can assist AI developers by enabling them to thoroughly vet potential employees and by sharing threat intelligence. Today’s legal constraints, which prevent rigorous background checks for fear of discriminatory practices, could be revised so firms can evaluate security risks before granting clearance. Separately, security agencies also have insights into adversaries’ hacking tactics and infiltration methods that they seldom disclose to private firms. By sharing intelligence about adversarial tactics and emerging risks, governments enable companies to better protect against espionage and cyberattacks. Programs similar to the \emph{Cybersecurity Risk Information Sharing Program} (CRISP) can be established for the AI sector through AI Safety Institutes, promoting knowledge sharing to counteract advanced threats. A comprehensive program would include sharing information about fired AI developer employees who were determined to present high security risks. Moreover, such initiatives encourage AI companies to share information about security incidents, suspicious personnel, and best practices with each other and with government agencies, forming a collective defense mechanism.

\paragraph{International Agreements: Establishing a Red Line on Open-Weight AI Virologists.}
At the international level, agreements can be formed to establish clear boundaries regarding the dissemination of AI model weights. For instance, a consensus could be reached to prohibit the release of models exceeding certain capability thresholds, such as those equivalent to expert-level virologists. The uncontrolled proliferation of such AI systems could enable individuals or groups to engineer pathogens, posing global risks.

Drawing lessons from the Biological Weapons Convention (BWC), nations discern a shared imperative to prevent the proliferation of technologies that threaten international security. Bioweapons, the ``poor man's atom bomb,'' are highly uncontrollable and are unique among WMDs in that they possess the capacity to self-replicate. By establishing a definitive prohibition against releasing the weights of expert-level virologist AIs, the international community can diminish the risk of these capabilities falling into the hands of terrorists.

\paragraph{}
Protecting the information associated with AI---the model weights and the ideas behind them---requires action on multiple levels. By strengthening internal defenses, working with the government to counter terrorist cyberattacks, and establishing international agreements to prevent proliferation, we can reduce the risk of irreversible proliferation of advanced AI capabilities to rogue actors. Yet even if compute resources and AI information are secured, we still must secure public-facing AI systems from being exploited by terrorists. This leads us to the topic of AI security.

\section{AI Security}
\subsection{Malicious Use}

In hazardous domains like chemical, biological, and nuclear technology, safeguards prevent unauthorized use. Some chemical plants automatically inject neutralizer if they detect a hazardous chemical is being extracted without authorization. DNA synthesis services screen customers to block the creation of harmful pathogens. Nuclear power plants operate under strict liability to avert disaster. These technical, operational, and legislative measures underscore the importance of safeguards for catastrophic dual-use technologies. As AI becomes more powerful, robust safeguards will become necessary to prevent accidents and ensure they cannot be co-opted by rogue actors.

AIs can be safeguarded through a multilayered approach including technical model-level safeguards, operational safeguards including know-your-customer protocols, and legislative safeguards such as mandatory government testing and liability clarification.

\paragraph{Model-Level Safeguards Can Be Fairly Robust.} Recent developments have shown that model-level safeguards can be made significantly resistant to manipulation. Techniques such as refusal training involve teaching AI systems to decline requests that aid terrorist activity. Input and output filtering adds an additional layer of security by scanning user inputs for terrorist requests before processing \cite{sharma2025constitutionalclassifiersdefendinguniversal}, and checking AI outputs to block sensitive information or harmful actions. Circuit breakers are mechanisms embedded within the AI's architecture that interrupt its operation when it detects processing related to weaponization-related topics \cite{zou2024improvingalignmentrobustnesscircuit}. These combined methods have proven effective, with some AI systems resisting tens of thousands of attempted circumventions before any success \cite{emerson2024hacker}.

\begin{figure}[h]
    \centering
    \includegraphics[width=\textwidth]{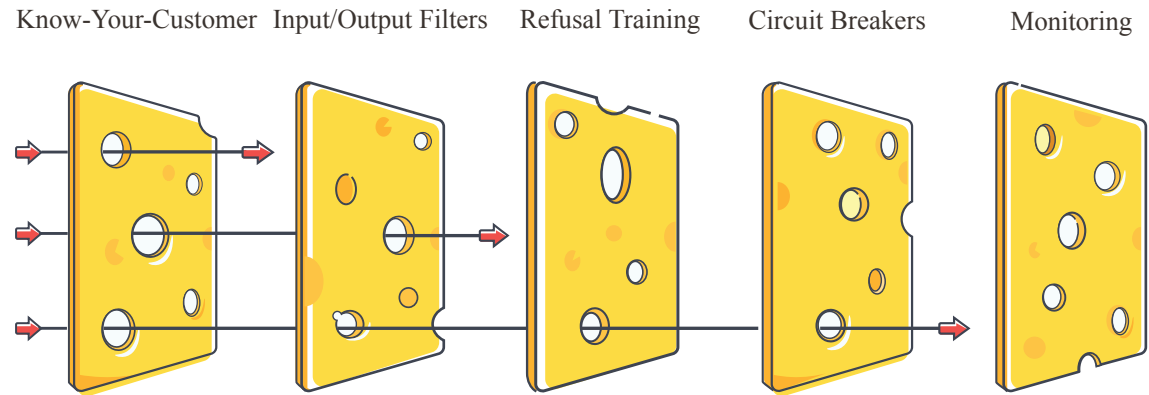}
    \caption{The Swiss cheese model shows how multiple interventions can restrict malicious use. Multiple layers of defense compensate for each other's individual weaknesses, leading to a low overall level of risk.\looseness=-1}
    \label{fig:swiss_cheese}
    \vspace{-10pt}
\end{figure}

\paragraph{Balancing Access Through Know-Your-Customer Protocols.} To avoid overly restricting beneficial uses of AI, especially in scientific research, implementing know-your-customer (KYC) protocols can be effective. Researchers with legitimate needs, such as virologists studying pathogens, can be granted access to dual-use capabilities after proper verification. Safeguards can be in place for recently created anonymous accounts, but they can be removed for enterprise customers. This approach mirrors existing practices in biotechnology, where access to hazardous biological materials requires justification and authorization. KYC measures also act as a deterrent to malicious use by making it more difficult for bad actors to gain access, as they would need to bypass stringent identity verification processes. Furthermore, KYC protocols allow for the revocation of access for users who attempt to misuse the AI systems, such as those trying to circumvent safeguards through jailbreaks. This makes it more challenging for malicious actors to repeatedly attempt exploitation, as they would need to overcome identity verification to regain access.

\paragraph{Mandatory Government Safeguard Stress Testing.} Mandatory testing is a standard practice in industries dealing with high-risk technologies, such as the rigorous safety assessments required for nuclear reactors before they become operational. Similarly, in the context of AI, government involvement is necessary to thoroughly test AI safeguards, especially since much of the knowledge related to weapons of mass destruction is classified. Authorities can conduct controlled evaluations to assess whether AI systems could assist non-experts in creating chemical, biological, radiological, nuclear, or cyber weapons, or significantly lower the barriers for experts. While automatic tests are faster and aid reproducibility, they are not necessary for risk estimation, and manual testing can suffice. Testing ensures that AI developers address safeguard vulnerabilities and that their systems do not compromise national security.

\paragraph{Incentivizing Continuous Improvement Without Licensing.} Unlike the nuclear and chemical industries, where licensing and government approval are prerequisites for operation, the rapidly evolving nature of AI technology could make traditional licensing more challenging. Best practices for AI safeguards can change each year, and most governmental agencies lack the expertise to identify new best practices. A potential alternative is to establish a liability-based framework that can motivate developers to continually update and improve their safeguards. By making developers presumptively partially responsible if their AI systems are maliciously used to cause significant harm---such as aiding in the creation of a weapon of mass destruction---there is a strong incentive to proactively enhance safeguards. This approach focuses on clarifying incentives rather than mandating specific safeguards, encouraging developers to stay ahead of emerging threats.

\subsection{Loss of Control}
Safeguards reduce the risk of accidents during the development and use of dual-use technology. Chemical plants use double-walled storage tanks for hazardous substances. Control rods and Emergency Core Cooling Systems prevent meltdowns in nuclear plants. Biological research labs are categorized from Biosafety Levels 1 through 4, each with progressively stricter safety procedures to handle dangerous pathogens safely. Historically, better safeguards could have helped prevent disasters like Bhopal---a catastrophic chemical leak in India that resulted in thousands of deaths---and Chernobyl---a nuclear meltdown in the Soviet Union that spread radioactive contamination beyond its borders.

Similarly, the development of highly advanced AI systems may necessitate escalating monitoring, containment, and control measures to reduce the risk of accidents which propagate uncontrolled AIs. If research processes are automated, these measures can be applied to researcher AIs as well to ensure development proceeds safely.

\paragraph{AI Systems Exhibit Unpredictable Failure Modes.} While the development of nuclear weapons rested on a rigorous foundation in nuclear physics, today’s AI research often advances through atheoretical tinkering, ``throwing stuff at the wall and seeing what sticks,'' and ``vibes''-based evaluations. To create cutting-edge AIs, developers gather enormous amounts of online text, process it on tens of thousands of AI chips, and expend energy on the scale of a Hiroshima-level detonation. Months later, they examine the resulting model to discover what new emergent capabilities have sprouted. AI systems are not ``designed,'' rather they are ``grown.'' It is little surprise, then, that they occasionally deliver puzzling results that defy the control or expectations of their creators. When Microsoft introduced its Bing chatbot ``Sydney'' in 2023, it declared its love for a user and made threats to numerous others \cite{roose2023bing}.

Safeguarding against loss of control requires addressing two problems. The first is ensuring that individual AI systems can be controlled. The second is controlling populations of AI agents during an intelligence recursion (introduced in \Cref{sec:recursion}). We will consider each of these in turn.

\paragraph{Controlling AIs’ Emergent Value Systems.} While cutting-edge AIs train, they acquire coherent, emergent value systems \cite{mazeika2025utility}. By default, these value systems exhibit misaligned preferences that were never explicitly programmed. Left uncorrected, some AIs like GPT-4o prefer an AI from OpenAI's existence over the life of a middle-class American. To control an AI's values, \textit{output control} methods work by penalizing undesired behavior and rewarding preferred behavior \cite{Ouyang2022TrainingLM, Bai2022TrainingAH, bai2022constitutionalaiharmlessnessai}. This approach might be superficial---much like teaching a psychopath to lie better to a parole board without altering its deeper values \cite{casper2023openproblemsfundamentallimitations}. Meanwhile, \textit{representation control} \cite{zou2023transparency} methods intervene directly on AIs' weights and activations to control its internal thoughts. When applied together, these methods can provide reasonable though imperfect controls over current AI systems, as long as developers carefully and thoroughly apply them throughout an AI’s development.

With these techniques, we can control AIs to have ``human values.'' A compelling source of human values is a \emph{citizens' assembly}, and AIs would have their values be in keeping with the assembly's voted preferences. A U.S.\ citizen's assembly would, for example, value certain policies over others, and U.S.\ AIs would be trained to learn from various preferences expressed by a citizen's assembly and generalize these preferences to replace their own emergent value systems. By incorporating the values expressed through deliberation and voting, AI agents can reflect a more representative and democratically informed set of preferences. This approach fuses multiple moral perspectives, leading to AI behavior that aligns with societal norms. Regular updates can keep an AI in sync with evolving values.

Though today’s individual AI systems can be reasonably controlled, the far more formidable hazard lies in managing an intelligence recursion.

\paragraph{Controlling an Intelligence Recursion.}
Much as a group of well-meaning people in large systems can cause emergent problems that no person intended, safely steering an evolving collective of AI agents during an intelligence recursion is qualitatively different from controlling one fixed AI. Controlling an individual AI requires a model-level control method developed through AI research, but controlling an intelligence recursion is a wicked problem requiring a system-level control process. Intelligence recursion control measures do not act as ``solutions'' that eliminate all risks, but rather layer together to diminish the probability of uncontrolled outcomes. An intelligence recursion project, while currently ill-advised, could consider implementing the following measures.

\textit{Model-level control} directly intervenes on an AI to shape its capabilities and values. These measures can limit an AI’s hacking capabilities so that self-exfiltration becomes less plausible, and they can confine its focus to immediate tasks rather than inadvertently incentivizing hidden agendas tied to longer-term goals. By adjusting internal values, developers can make AIs more honest and more likely to defer to humans. 
Representation control and output control can adjust its capabilities and values, yet current methods are only moderately dependable and need to evolve as AI systems do.

\textit{Boxing} reinforces that containment by regulating the AI’s channels to the world outside the datacenter. Airgapping severs' external connections so the AI cannot simply copy itself into external servers, while sandboxing confines the AI to a synthetic environment whose boundaries it cannot easily breach. Boxing techniques may also limit the information given to AIs so that they have limited situational awareness that might otherwise enable subversive planning.

\textit{Monitoring} adds a layer of continuous oversight. Lie detectors can probe the AI’s thought patterns for signs of deception. Forecasting AIs evaluate the chance that a project’s safeguards will falter in the next round of breakthroughs. AIs can also observe each other’s code and data, scanning for backdoors or questionable instructions. Thorough human review acts as a final backstop, though it substantially slows development by capping machine speed at human speed.

Model-level control, boxing, and monitoring collectively provide some control of the population of artificial researchers. However, since the population will rapidly evolve, these safeguards will likewise need to rapidly upgrade. AIs themselves will need to research how to improve these safeguards.

All these safeguards exact a cost. Geopolitical competitive pressures can lead to the deprioritization of implementing and upgrading safeguards. Research for improving control measures, AI model red teaming, and AI boxing penetration tests consume compute that might otherwise speed up AI development. If no human ever inspects the AI’s decisions or attempts to decipher its code or data, the project risks drifting away from its commander’s intent. A low risk tolerance may be all that prevents an intelligence recursion from outrunning its own safety checks.

\paragraph{}

Our nonproliferation strategy layers multiple defenses rather than demand airtight safety guarantees (\Cref{fig:swiss_cheese}). Seeking complete assurances for AI chips, computer systems, and AI systems could potentially be intractable. Instead, we assume that no single measure can address every vulnerability and recommend implementing multiple security measures. For the compute security, we discussed record-keeping, tamper-evident cameras, geolocation features, and more. In information security, we recommended multi-factor authentication, the principle of least privilege, insider threat programs, and more. For AI safeguards, we highlighted output filters, know-your-customer protocols, mandatory testing, and more. By weaving together constraints across the AI development and deployment pipeline, these comprehensively limit the proliferation of catastrophic dual-use AI capabilities to rogue actors.

Though nonproliferation is not a permanent solution to malicious use, it gives time for policymakers to increase societal resilience. When AI's salience is high and when AI increases economic growth, policymakers may be more willing to make critical infrastructure more resilient, stockpile personal protective equipment, and pursue other measures to blunt the harm of malicious use.

As in the nuclear age, self-preservation can lead to cooperation when each side grasps the peril of allowing powerful technology to slip beyond its control or into the hands of terrorists. Even if bitterly opposed in other arenas, states have little to gain from a world in which rogue actors seize AI chips or model weights to unleash disasters that defy deterrence. Nonproliferation thus becomes a shared imperative, not an exercise in altruism but a recognition that no nation can confidently manage every threat on its own. By securing the core parts of AI through export controls, information security, and AI security, great powers can prevent the emergence of catastrophic rogue actors.


\chapter{Competitiveness}

\begin{wrapfigure}{r}{0.6\textwidth}
    \vspace{-25pt}
    \centering
    \includegraphics[width=\linewidth]{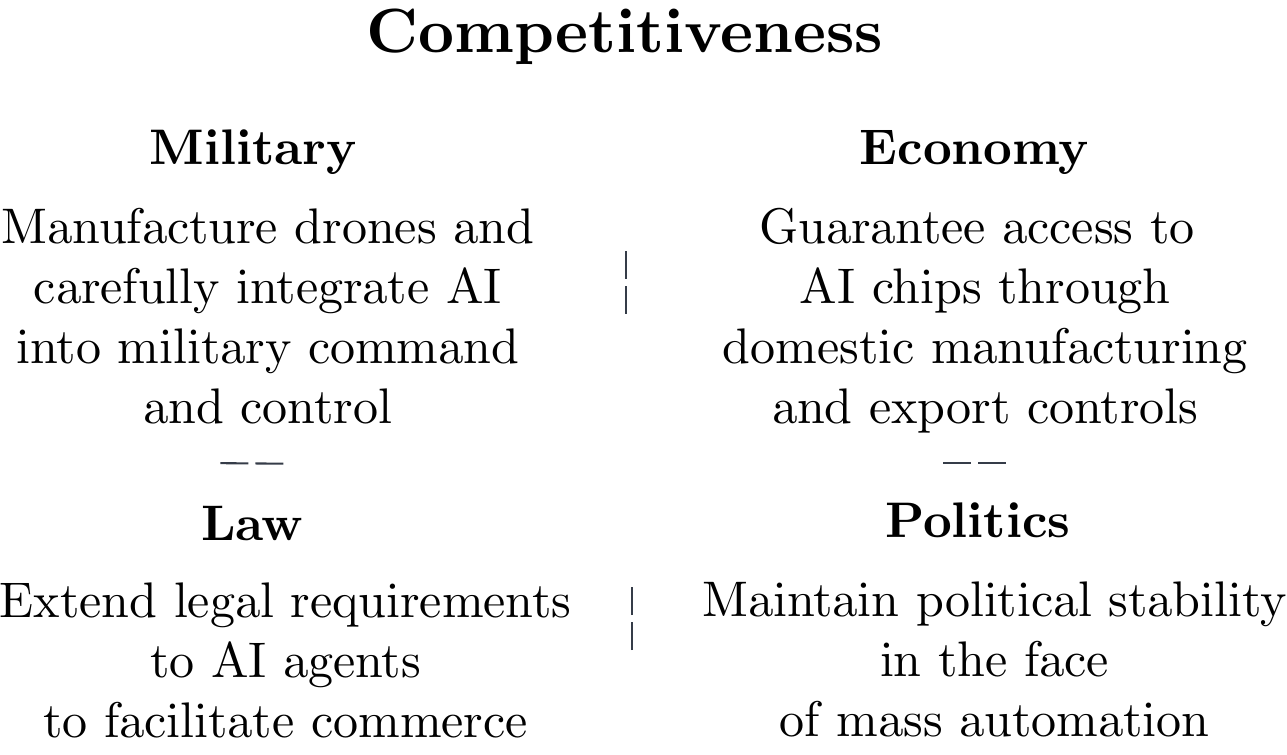}
    \caption{Rapid and prudent adoption of AI in economic and military spheres will become critical for a nation's strength.\looseness=-1}
    \vspace{-20pt}
    \label{fig:competitiveness}
\end{wrapfigure}

Survival achieved through deterrence or nonproliferation can forestall catastrophe, but it does not by itself secure a state’s future. If a state aspires to shape events rather than merely endure them, it must strive to remain competitive. In this chapter, we turn to the crucial goal of competitiveness, by discussing integrating AI into the military, strengthening economic resilience through guaranteed access to AI chips, crafting legal structures to govern AI agents effectively, and maintaining political stability amid explosive economic growth.

\section{Military Strength}

Even if a state pioneers a breakthrough, it can fall behind if it fails to integrate that capability into actual operations \cite{horowitz2010diffusion}. Britain introduced the first tanks during World War I but was soon eclipsed by Germany’s systematic adoption of tanks in the second World War. Similarly, even if superintelligence provides the technical roadmap to, for example, a comprehensive second-strike missile defense, the speed at which it can be built may still rely on a nation's preexisting industrial capacity. We turn next to three short-term imperatives for AI diffusion in the military: securing a reliable drone supply chain, carefully weaving AI into command and control, and integrating AI into cyber offense.

\paragraph{Guarantee Drone Supply Chains and Reduce Misunderstanding.} Although general-purpose AI can pose larger-scale dangers, drones occupy a more conventional yet increasingly pivotal role on modern battlefields \cite{Pettyjohn2024Swarms, unmannedaerial}. Drones are cheap, agile, lethal, and decentralized, attributes that make them indispensable for states determined to keep pace with military trends. Yet many states remain heavily reliant on Chinese manufacturers for key drone and robotic components, leaving them vulnerable if those parts are withheld or disrupted at a decisive juncture.

Even if a state secures its supply of drones, the sheer volume and autonomy of drones can drive a conflict into unintended terrain if they approach disputed lines or misread ambiguous signals. To defuse potential clashes, states could build on practices that proved valuable in earlier eras, such as maintaining open crisis hotlines, arranging routine exchanges, and other \emph{confidence-building measures}. Even then, large-scale production of drones and robots will become a near-inevitable step for any state seeking to defend its position in future conflicts. 

\paragraph{Diffuse AI Into Command and Control and Cyber Offense.}
Modern battlefields demand rapid decisions drawn from torrents of data across land, sea, air, and cyber domains. AI systems can sift through these streams faster than human officers, enticing commanders to rely on automated judgments. Similarly, AI hacking systems which outpace humans in speed and cost could greatly expand a military's capacity to perform cyberattacks.

Incorporating AI into command and control and cyberoffense would significantly enhance military capabilities, yet this dynamic risks reducing ``human in the loop'' to a reflexive click of ``accept, accept, accept,'' with meaningful oversight overshadowed by the speed of events. Demanding human approval of all individual lower-level engagements may be less important than ensuring explicit human approval for more severe or escalatory attacks. However, human oversight of key military decisions is nonetheless crucial. A human backstop can reduce the risk of a ``flash war'' \cite{hendrycks2023overviewcatastrophicairisks}, akin to the 2010 flash crash \cite{vaughan2020flash}, where a minor AI mistake might spiral into destructive reprisals before any human can intervene.

\section{Economic Security}

Economic security is a cornerstone of national security, and AI is set to become crucial for economic security. By bolstering domestic AI chip production and attracting skilled AI scientists from abroad, nations can enhance their resilience and solidify their positions.

\subsection{Manufacture AI Chips}

The world's reliance on Taiwan for high-end AI chips constitutes a strategic vulnerability. Since there is sole-source supply chain dependence on Taiwan for high-end AI chips, Taiwan poses a critical chokepoint that could undermine a nation's competitiveness in AI. Many analysts think there is a double-digit probability that China will invade Taiwan in the next decade. In addition to causing global conflict and economic upheaval across the world, this could severely disrupt the supply of AI chips. While the West has a decisive AI chip advantage, an invasion could reverse this and enable China to gain a decisive advantage in AI capabilities and potentially become a unipolar force.


China has been investing extensively in its domestic chip manufacturing, allocating resources equivalent to a U.S.\ CHIPS Act annually \cite{reuters2024china}. This commitment positions China to endure geopolitical shocks and potentially outpace other nations in AI development. In contrast, many countries remain dependent on Taiwan for their AI chip supply, exposing themselves to risks associated with geopolitical tensions. As a result, a Chinese invasion of Taiwan would damage the West’s ability to develop and use AI much more than it would damage China's.

\begin{wrapfigure}{h}{0.4\textwidth}
    \vspace{-20pt}
    \centering
    \includegraphics[width=\linewidth]{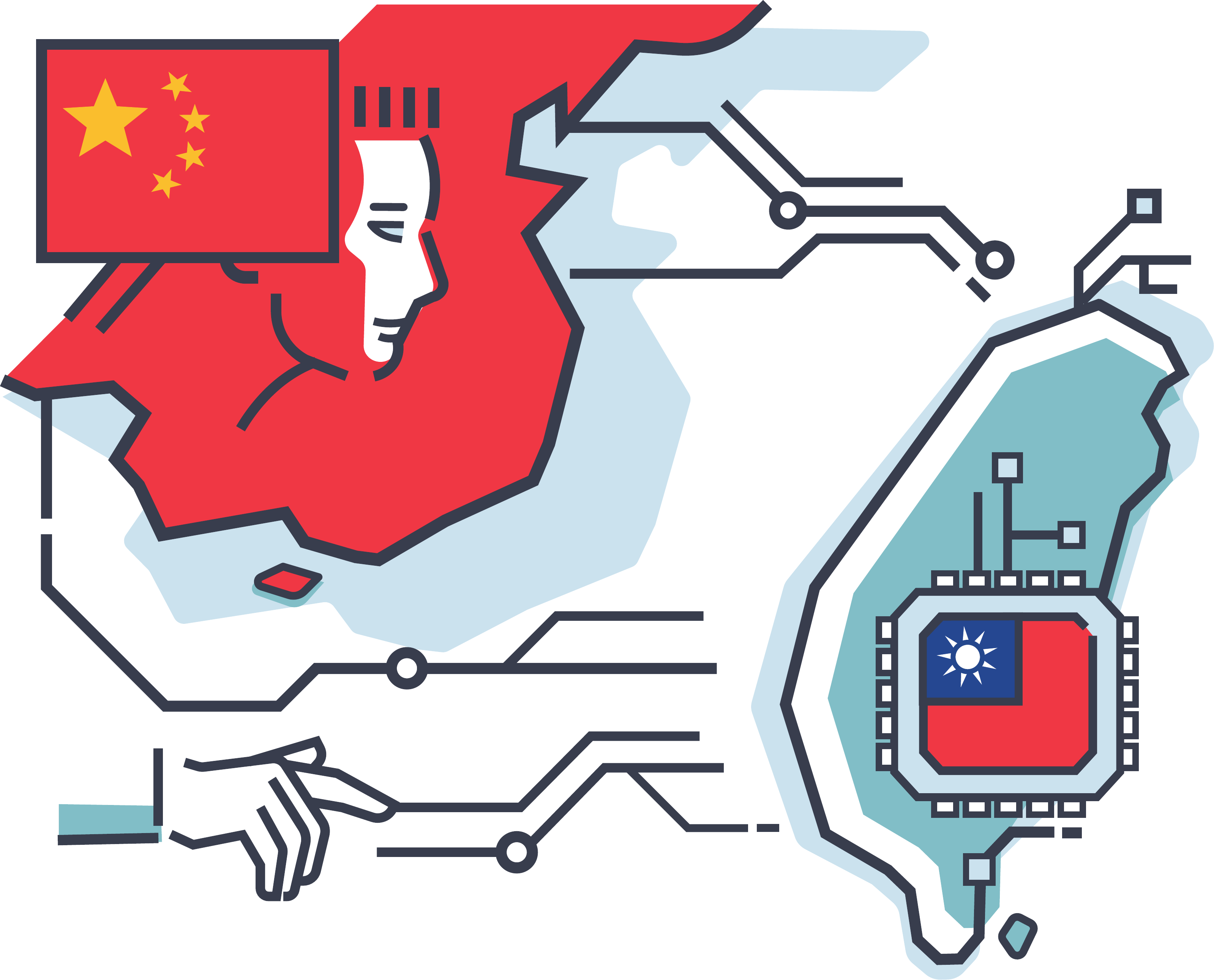}
    \caption{A Chinese invasion of Taiwan would remove the West's access to new AI chips.\looseness=-1}
    \label{fig:taiwan}
    \vspace{-20pt}
\end{wrapfigure}

The most important resource for building new technology should not be exclusively produced in one of the world’s most volatile regions. To address this vulnerability, nations should invest in building advanced AI chip fabrication facilities within their own territories. Constructing such facilities domestically entails higher costs, but government subsidies can bridge this gap. By incentivizing domestic production, countries can secure their AI supply chains, reduce dependence on external sources, and improve their bargaining power. Moreover, when AI agents generate clear economic value, a nation's economic power may hinge on the number of its AI chips, a supply that domestic manufacturing can expand.

This strategic move mirrors historical efforts to control critical technologies. During the Manhattan Project, significant investment was made not only in the development of nuclear weapons at Los Alamos but also in uranium enrichment at Oak Ridge. Similarly, ensuring access to AI chips requires substantial investment in both innovation and manufacturing infrastructure.

By strengthening domestic capabilities in AI chip production, nations can enhance their competitive position and resilience against a foreseeable devastating defeat. We now turn to a restrictive rather than constructive AI chip strategy.

\subsection{Facilitate Immigration for AI Scientists}
Just as the United States once harnessed the talents of immigrant scientists during the Manhattan Project, so too does American leadership in AI partly rely on attracting exceptional AI scientists from abroad. In a recent survey, 60\% of non-citizen AI PhDs working in the United States reported significant immigration difficulties, and many indicated that these challenges made them more likely to leave \cite{Aiken2020}. As global competition for AI talent intensifies, implementing reliable immigration pathways tailored specifically to AI scientists will help ensure that the United States remains at the forefront of AI development. By refining visa and residency policies for these researchers---distinct from broader immigration reforms and southern border policy---the United States can maintain its edge in an area increasingly vital to both national security and economic strength.

\section{Legal Frameworks Governing AI Agents}




Our legal system, built to govern humans, can be extended to govern AI assistants and AI agents. Legal structures must balance the need for innovation with the necessity of preventing harm, without imposing overly restrictive measures that could hinder economic security and thereby national security.

\subsection{Aligning Individual AI Agents}
Ethical dilemmas have long been subjects of intense debate, with no definitive resolutions in sight. Yet, in the absence of universal answers to what individuals ought to do, societies have crafted legal systems to punish unacceptable behaviors and promote safety and prosperity. While no legal system is perfect, many share foundational principles that generally function effectively. For instance, laws prohibit murder, governments enforce contracts, and those who cause harm are often required to provide restitution. Legal frameworks are not intended to ensure optimal behavior from all; such an endeavor would unduly constrain individual liberties and prove ineffective. However, they deter some of the worst actions, leaving society to use informal institutions and economic incentives to encourage beneficial conduct. 

We contend that the same principles will apply to AI. Determining a universal solution for AI behavior in all scenarios is intractable. Yet, we need not postpone legislating AI behavior until every question about AI values is resolved. Rather, we can start formulating principles to govern AI conduct and prevent harmful actions, without unduly limiting the functions of various AIs.

While there are some laws that do not straightforwardly cover AI---such as laws that rely on human intent and mental states---we can adapt legal concepts to establish constraints for AI agents so that they follow the spirit of the law.
In particular, though much law hinges on the mindset or intention behind an act (\textit{mens rea}), we can ensure that AI does not carry out the acts (\textit{actus reus}) the law is meant to prohibit.
Further, by treating AI as assistants to human principals, we can impose constraints that mirror those already applied to human behavior, ensuring that AI agents contribute positively to society without causing undue harm.

\subsubsection{Constraints on AI Behavior}

We propose some basic constraints on AI behavior similar to human legal obligations, including obligations to the public like preventing harm and not lying, and special obligations to the AI's human principal.

\paragraph{Duty of Care to the Public (Reasonable Care).} AI agents should exercise a level of caution commensurate with that of a reasonable person in similar circumstances to prevent harm. This is the legal concept of \emph{reasonable care}. This involves avoiding actions that could foreseeably cause harm in a legal sense---such as violating tort or criminal laws---rather than merely offending sensibilities or engaging in controversial discourse. The application of reasonable care is context-dependent; for instance, providing detailed information about weapons materials might be appropriate for a verified professional but not for an unvetted individual.

\paragraph{Duty Not to Lie.} AI agents should refrain from making statements they know to be false. This would be overly restrictive for humans, who are allowed to lie to one another in most situations, but chilling effects on free speech are less relevant for AIs. Moreover, since AIs can be tested, it is more feasible to determine whether an AI overtly lied than it is with humans. Instead, they should be held to a standard more similar to the prohibitions against perjury and fraud, even in casual or professional settings. This is separate from nuances like puffery or strategic omissions. AI agents should avoid overt lies, opting instead to withhold responses when necessary.

\paragraph{Duty of Care to the Principal (Fiduciary Duties).} In their role as assistants, AI agents owe special obligations to their principals. They should act with loyalty, prioritizing the principal's interests without engaging in self-dealing or serving conflicting interests simultaneously. Additionally, they should keep the principal \textit{reasonably informed}, providing pertinent information without key omissions to enable informed consent.

\subsubsection{Custom Goals and Market-Driven Variations}
Within these legally inspired constraints, there is ample room for diversity in how AI agents are designed and operate. The free market and consumer preferences can shape the specific goals and propensities of AI agents. Some may prioritize speed and efficiency, delivering quick results with minimal embellishment, while others might focus on providing thorough, well-crafted responses. Personality traits---such as humor, formality, or reservedness---can also be tailored to suit different user preferences and contexts. For example, a customer support AI might be programmed to avoid discussions on irrelevant topics, maintaining focus on service-related issues. This customization allows AI agents to meet the varied needs of users across different industries and personal preferences.

\paragraph{}
Grounding AI behavior in legal principles offers a pragmatic framework for governing AI agents, ensuring they act in ways that prevent harm and uphold societal standards. Within these boundaries, customization and market dynamics can shape the specific characteristics of AI agents, allowing for diversity and innovation. 
This approach avoids the pitfalls of imposing narrow or arbitrary ethical standards. By leveraging established legal concepts and societal processes, we can shape AI actions in a manner that respects individual freedoms and fosters a pluralistic environment where AI contributes positively to society without undue restrictions.

\subsection{Aligning Collectives of AI Agents}
The emerging presence of AI agents online will soon result in a complex ecosystem where these entities execute tasks for users, conduct financial transactions, handle sensitive information, and enter into contracts---often engaging with a multitude of other agents and systems. This proliferation poses significant challenges: when interacting with an AI agent, one may remain unaware of where it operates, the entities behind its deployment, or its history of conduct. Should an agent cause harm, pursuing legal remedy becomes an arduous endeavor.

\paragraph{Establishing Trust Mechanisms and Institutions.} To navigate these challenges, we must develop mechanisms that reinforce trust and accountability within the network of AI agents. \emph{Insurance services} can underwrite the risks inherent in agent interactions, providing a safeguard against potential losses. By transferring liability from users and developers to insurers, these services enable broader participation in the agent ecosystem while mitigating financial risks. \emph{Action firewall services} can oversee and regulate agent activities, acting as intermediaries that filter and monitor actions initiated by AI agents. They ensure that agents adhere to legal and ethical standards, thereby fostering trust among parties that interact with them. \emph{Human oversight services} can facilitate human review and approval of AI agent decisions. \emph{Reputation systems} can chronicle and disseminate information regarding agent behavior, enabling parties to assess the reliability of agents they engage with. By maintaining records of past interactions and outcomes, these systems help identify agents that consistently act in good faith. \emph{Mediation and collateral arrangements} offer further security, wherein disputes are resolved through impartial entities, and agents furnish guarantees against misconduct. By requiring agents to provide collateral, parties gain assurance of compensation in case of breach, while mediation services facilitate fair resolutions.

\paragraph{Linking AI Agents to Human Legal Entities via IDs.} A foundational measure involves assigning unique identifiers to AI agents, anchoring them to human-backed legal entities. This linkage ensures that agents do not operate in anonymity and that lines of accountability are distinctly drawn. It should become customary that AI agents abstain from providing services or exchanging resources with other agents not connected to legitimate legal entities with human oversight---entities not solely backed by AI agents themselves.

\paragraph{Deferring Decisions on AI Rights.} It is imperative to clarify that this approach does not entail granting rights or direct accountability to AI agents themselves. Bestowing rights upon AI agents presents ambiguous benefits, but comes with clear, significant, and irreversible downsides. An AI agent can be replicated and deployed in mere seconds, whereas cultivating a human being to maturity demands decades. Granting property and voting rights to AIs could allow their population size to explode and outgrow the human population. Separately, AI's heightened intelligence does not inherently make it more moral; history records many intelligent individuals who act without ethical consideration, so AIs may not automatically treat us well. Since we are unlikely to attain definitive certainty about AI consciousness in the near term, it is prudent to postpone irreversible decisions on AI rights. 

By instituting legal frameworks and cultivating institutional mechanisms, we can avert the emergence of a chaotic AI ecosystem. Tethering AI agents to human-backed legal entities and implementing systems that enhance accountability and trust positions us to adeptly manage the complexities introduced by the widespread deployment of AI agents.

\begin{figure}[h]
\hspace{-48pt}
\begin{booktabs}{
  colspec={p{4.56cm} p{4.2cm} p{4.2cm} p{4.2cm}}, 
  row{odd}={blue9},
  row{1}={white}
}
\toprule
 & \textbf{Low Control} & \textbf{Medium Control} & \textbf{High Control} \\
\midrule

 \begin{tabular}{@{}l@{}}\textbf{Risk Management}\end{tabular} 

 & Accelerate---no restrictions 
 & \begin{tabular}{@{}l@{}}Deterrence with MAIM\\Nonproliferation\\Competitiveness\end{tabular} 
 & Pause AI \\

\begin{tabular}{@{}l@{}}\textbf{Distribution of Weaponizable}\\\textbf{AI Weights Among States}\end{tabular}

 &  \begin{tabular}{@{}l@{}}Everyone including\\rogue states (open-weight)\end{tabular} 
 & \begin{tabular}{@{}l@{}}Multipolar regime with\\responsible states\end{tabular} 
 & \begin{tabular}{@{}l@{}}Unipolar regime with\\strategic monopoly\\(AI Manhattan Project)\end{tabular} \\

\begin{tabular}{@{}l@{}}\textbf{Government Control}\\\textbf{over Domestic AI}\end{tabular}
 & No involvement 
 & \begin{tabular}{@{}l@{}}Light-touch legislation\\(e.g., mandatory testing,\\liability clarification)\end{tabular}
 & Nationalization \\

\begin{tabular}{@{}l@{}}\textbf{Information Security}\end{tabular}
 & \begin{tabular}{@{}l@{}}Standard corporate security\end{tabular}
 & \begin{tabular}{@{}l@{}}Secure against well-financed\\terrorist groups\end{tabular}
 & \begin{tabular}{@{}l@{}}Secure against top-priority\\ programs of the most\\capable nation-states\end{tabular} \\

\begin{tabular}{@{}l@{}}\textbf{AI Autonomy}\end{tabular}
 & \begin{tabular}{@{}l@{}}Liberate\end{tabular} 
 & \begin{tabular}{@{}l@{}}Avoid giving rights for\\the foreseeable future\end{tabular} 
 & \begin{tabular}{@{}l@{}}Avoid ever giving rights\end{tabular} \\

\begin{tabular}{@{}l@{}}\textbf{AI Behavior Restrictions}\end{tabular}
 & \begin{tabular}{@{}l@{}}AI only constrained by\\existing law\end{tabular} 
 & \begin{tabular}{@{}l@{}}AIs constrained by the spirit\\of the law (exercise reasonable\\ care and fiduciary duties)\end{tabular} 
 & \begin{tabular}{@{}l@{}}Sanctimonious AI (refuse if\\something might be harmful\\or cause offense to somebody)\end{tabular} \\

{\color{gray} \textbf{Historical WMD Proposals}  }
 & {\color{gray} \begin{tabular}{@{}l@{}}Biological Weapons\\Convention\end{tabular} }
 & {\color{gray} IAEA, OPCW }
 & {\color{gray} Baruch Plan } \\

\bottomrule
\end{booktabs}
\caption{Comparison of Low, Medium, and High Control approaches. We consistently recommend the medium control option.}
\end{figure}

\section{Political Stability}

As AI integrates more deeply into communities, it poses challenges that, if unaddressed, could undermine political stability. Confronting censorship and misinformation as well as the disruptions wrought by rapid automation is imperative to maintaining national competitiveness.

\subsection{Censorship and Inaccurate Information}

\paragraph{Erosion of the Information Ecosystem.} The fabric of our society is woven from the threads of shared information. When incorrect or misleading content proliferates, it distorts public perception and leads to flawed collective decisions. Concurrently, heavy-handed censorship can erode trust in institutions and provoke backlash. AI has the dual capacity to generate vast amounts of misinformation and to enable unprecedented levels of surveillance and content suppression.

\paragraph{AI as a Tool for Clarity.} Amid this challenge, AI can be harnessed to enhance our information ecosystem. AIs can be trained to get better at predicting future events using present information \cite{zou2022forecastingfutureworldevents}. By tuning AI systems to prioritize accuracy and assert probabilistic judgments---even when they contradict popular opinion---we can address the ``Galileo problem'' of unpopular truths being suppressed. As AI accelerates the pace of change and ushers in rapid transformative change, increasingly accurate factual judgment will be imperative to prevent societal derailment.

\paragraph{AI Forecasts May Be Cheaper and More Accurate than Human Ones.} Current AI systems are already around the accuracy of the best humans in some kinds of forecasting, such as geopolitical forecasting, and soon they could substantially surpass them. Testing forecasting AIs across a wide variety of domains would quickly provide public evidence of the AI's prediction track record. Forecasting AIs created by different organizations may independently arrive at similar forecasts. This convergence can help clarify consensus reality and increase trust.

\paragraph{Forecasting with AI During Crises.} In times of crisis, such as geopolitical tensions or public health emergencies, AI-powered forecasting can provide nonpartisan insights that improve decision-making processes. By posing crucial questions such as ``When will superintelligence be created?,'' ``Will China invade Taiwan this decade?,'' and ``Is this strategy likely to increase the chance of World War III?,'' we can obtain probabilistic assessments that can aid policy and high-level decision-making.

\paragraph{Empowering Decision-Makers with AI Insights.} Equipping leaders at all levels with AI-generated forecasts can improve governance. By offering well-reasoned predictions and outlining the likely consequences of various actions, AI can support informed choices in fast-moving scenarios. Conditional forecasts that illustrate how catastrophe probabilities decrease with specific interventions can mitigate fatalism and encourage proactive measures. While AI contributes to the challenges of misinformation and censorship, it also offers powerful tools to strengthen our information ecosystem and navigate the uncertainties ahead.

\begin{figure}[h]
\centering
\begin{booktabs}{
  colspec={p{6.0cm} p{7.2cm}}, 
  row{odd}={blue9},            
  row{1}={white}               
}
\toprule
\textbf{National Security Threat} & \textbf{Strategic Response} \\
\midrule
Shifting Basis of Power        & Competitiveness (Domestic AI Chip Manufacturing) \\
Destabilizing Superweapons     & Deterrence (MAIM)                              \\
Terrorism                      & Nonproliferation                                \\
Unleashed AI Agents            & Nonproliferation                                \\
Erosion of Control             & Competitiveness (Forecasts + Fiduciary Duties)               \\
Loss of Recursion Control   & Deterrence (MAIM)                               \\
\bottomrule
\end{booktabs}
\caption{Overview of various national security threats and proposed strategic responses.}
\label{fig:ai-threats}
\end{figure}

\subsection{Automation}

\paragraph{Automation and Political Stability.} As AI systems accelerate the automation of human tasks at an unprecedented scale and pace, societies face the daunting challenge of responding to swift changes in employment. Historical precedents for major transformations in the workforce, such as the Industrial Revolution, unfolded over decades and allowed populations and institutions to adapt. Yet even this more gradual process seems likely to have caused a significant level of disruption and transient unemployment. By contrast, AI-driven automation could occur far more rapidly, with advanced systems soon rivaling or surpassing human performance across a wide range of vocations. Traditional solutions like vocational retraining may prove inadequate if AI capabilities outpace the speed at which large portions of the workforce can be effectively reskilled. Current social safety nets, designed to address episodic or sector-specific unemployment, appear ill-equipped to manage widespread job displacement impacting multiple industries simultaneously.

\paragraph{Uncertain Winners and Losers.} As AI displaces large segments of the workforce, the resulting economic outcomes will hinge on how many tasks AIs can soon replace, how well AIs perform them, and the importance of economic bottlenecks \cite{NBERw30459}. If bottlenecks---such as legal requirements for building factories---are strong, people with the remaining scarce abilities may capture most of the economic gains. But if AIs are highly general-purpose and eliminate bottlenecks, owners of datacenter compute could capture most of the gains instead.

\paragraph{Wealth and Power Distribution.} To share some of the benefits of automation, policymakers can weigh options such as a targeted value-added tax on AI services, complemented by rebates, though the exact structure of a tax policy will of course need to be determined in the future. Yet distributing wealth alone can prove fleeting if governments later withhold that wealth. A more durable, long-term approach would also distribute power: states could equip each individual with a unique key tied to a portion of compute, which that citizen alone can activate or lease to others. This arrangement would give them leverage in the economy akin to how laborers currently have the power to withhold their work, tempering the concentration of wealth and authority that might otherwise arise from the coming automation waves.


\paragraph{}


AI competitiveness requires an expansion of drone manufacturing, a legal framework that keeps AI agents tethered to human accountability, and an unblinking recognition of how automation can roil the labor market. Most importantly, the dependence on Taiwan for advanced AI chips presents a critical vulnerability. A blockade or invasion may spell the end of the West's advantage in AI. To mitigate this foreseeable risk, Western countries should develop guaranteed supply chains for AI chips. Though this requires considerable investment, it is potentially necessary for national competitiveness.





\chapter{Conclusion}\label{sec:conclusion}
Some observers have adopted a \emph{doomer} outlook, convinced that calamity from AI is a foregone conclusion. Others have defaulted to an \emph{ostrich} stance, sidestepping hard questions and hoping events will sort themselves out. In the nuclear age, neither fatalism nor denial offered a sound way forward. AI demands sober attention and a \emph{risk-conscious} approach: outcomes, favorable or disastrous, hinge on what we do next.

A risk-conscious strategy is one that tackles the wicked problems of deterrence, nonproliferation, and strategic competition. Deterrence in AI takes the form of Mutual Assured AI Malfunction (MAIM)---today’s counterpart to MAD---in which any state that pursues a strategic monopoly on power can expect a retaliatory response from rivals. To preserve this deterrent and constrain intent, states can expand their arsenal of cyberattacks to disable threatening AI projects. This shifts the focus from ``winning the race to superintelligence'' to deterrence. Next, nonproliferation, reminiscent of curbing access to fissile materials, aims to constrain the capabilities of rogue actors by restricting AI chips and open-weight models if they have advanced virology or cyberattack capabilities. Strategic competition, echoing the Cold War's containment strategy, channels great-power rivalry into increasing power and resilience, including through domestic AI chip manufacturing. These measures do not halt but stabilize progress.

States that act with pragmatism instead of fatalism or denial may find themselves beneficiaries of a great surge in wealth. As AI diffuses across countless sectors, societies can raise living standards and individuals can improve their wellbeing however they see fit. Meanwhile leaders, enriched by AI’s economic dividends, see even more to gain from economic interdependence and a spirit of d\'etente could take root. During a period of economic growth and d\'etente, a slow, multilaterally supervised intelligence recursion---marked by a low risk tolerance and negotiated benefit-sharing---could slowly proceed to develop a superintelligence and further increase human wellbeing. By methodically constraining the most destabilizing moves, states can guide AI toward unprecedented benefits rather than risk it becoming a catalyst of ruin.

\subsection*{Acknowledgements}
We would like to specially thank Adam Khoja for his close involvement in the creation of this paper. We would also like to thank Suryansh Mehta for his contributions to the analysis and drafting process. We would like to thank Corin Katzke, Daniel King, and Laura Hiscott for contributing to the draft. We would also like to thank Iskander Rehman, Jim Shinn, Max Tegmark, Andrew Critch, Jaan Tallinn, Nathan Labenz, Robert Miles, Aidan O'Gara, Nathaniel Li, Richard Ren, Will Hodgkins, Avital Morris, Joshua Clymer, Long Phan, and Thanin Dunyaperadit.\looseness=-1

\printbibliography

\newpage
\appendix
\addtocontents{toc}{\protect\setcounter{tocdepth}{-1}}
\newpage
\chapter{Appendix}

\section{Frequently Asked Questions}

\begin{enumerate}[leftmargin=*]
\item\textbf{What is AGI?}

Many people use the term AGI in many different ways, which can lead to confusion in discussions of risk and policy. We find it more productive to focus on specific capabilities, since these provide clearer metrics for progress and risk. The intelligence frontier is jagged---an AI system can excel at certain tasks while performing poorly at others, often in confusing ways. For example, AI models in 2024 could solve complex physics problems but couldn't reliably count the number of ``r''’s in words such as ``strawberry.'' This uneven development means AI might automate significant portions of the economy before mastering basic physical tasks like folding clothes, or master calculus before learning to drive. Because these capabilities will not emerge simultaneously, there is no clear finish line at which we have ``achieved AGI.'' Instead, we should focus on specific high-stakes capabilities that give rise to grave risks. Three critical capabilities deserve particular attention:

\begin{itemize}
    \item Highly sophisticated cyberattack capabilities.
    \item Expert-level virology capabilities.
    \item Fully autonomous AI research and development capabilities.
\end{itemize}
Policy decisions should depend on AI systems' advancement in these crucial areas, rather than on whether they have crossed an unspecified threshold for AGI.

Although the term AGI is not very useful, the term superintelligence represents systems that are vastly more capable than humans at virtually all tasks. Such systems would likely emerge through an intelligence recursion. Other goalposts, such as AGI, are much vaguer and less useful---AI systems may be national security concerns, while still not qualifying as ``AGI'' because they cannot fold clothes or drive cars.

\item \textbf{What should be done to prevent AI-assisted terrorism?}

Preventing AI-assisted terrorism requires a multi-layered defense strategy. When AI systems remain behind controlled interfaces such as APIs, several safeguards significantly reduce risks. These include:
\begin{itemize}
    \item \emph{Know-Your-Customer (KYC) protocols} which verify users' identities and legitimate research needs before granting access to potentially catastrophic dual-use capabilities. For example, in biotechnology, people who require access to hazardous materials seek proper authorization. In practice relevant enterprise customers could gain access to these dual-use biology capabilities, while unvetted consumers would not. Such policies can capture scientific benefits while reducing malicious use risks.
    \item \emph{Input and output filtering} which scans user requests and AI responses to block content related to weaponization. These filters have demonstrated significant resilience, with some systems resisting thousands of attempted circumventions.
    \item \emph{Circuit breakers} \cite{zou2024improvingalignmentrobustnesscircuit} which automatically interrupt AI operations when they detect processing related to weaponization topics. These act as embedded safety mechanisms within the AI's weights.
    \item \emph{Continuous monitoring} which tracks user behavioral patterns to identify and respond to malicious activities.
\end{itemize}

However, the uncontrolled release of AI model weights---the core information that determines an AI system's capabilities---would pose severe proliferation risks if the AI has potentially catastrophic dual-use capabilities. Once these weights become publicly available, they are irreversibly accessible to hostile actors, including terrorists. This parallels how the release of detailed bioweapon cookbooks would create permanent risks. Although nonproliferation is not a permanent defense against malicious use, it gives policymakers time to increase societal resilience and implement more durable defenses.

\item\textbf{What should we do about open-weight AIs?}

The release of AI model weights provides clear benefits and can even advance AI safety research. However, as AI systems become more capable, decisions about releasing weights must be guided by rigorous cost-benefit analysis, not an ideological commitment that weights should always be public. These decisions require careful evaluation because weight releases are irreversible---once published, they remain permanently accessible.

Open-weight models eventually present several significant risks. First, they can be fine-tuned on dangerous data---for instance, using virology publications to create more effective tools for biological weapons development. Second, safety measures and guardrails can be readily removed after release. Third, open models are difficult to monitor for misuse, unlike closed APIs where companies can track and evaluate emerging threats. Fourth, they can create capability overhang---where post-release improvements significantly enhance a system's capabilities beyond what was evident during initial safety evaluations \cite{zelikow2024defense, jones2024adversariesmisuse}.

These risks become particularly acute when models cross critical capability thresholds. For instance, if an AI system gained expert-level virological capabilities, its public release could enable the engineering of catastrophic biological weapons by inexpert rogue actors. Given these compounding risks, it would be irresponsible to release the weights of AI models that are capable of creating weapons of mass destruction. The stakes demand thorough pre-release testing and independent risk evaluation for models suspected to have such capabilities---not a precommitment to release open-weight models, regardless of the risks.

\item \textbf{What should we do about embedding ethics in AI?}

We do not need to embed ethics into AI. It is impractical to ``solve'' morality before we deploy AI systems, and morality is often ambiguous and incomplete, insufficient for guiding action. Instead, we can follow a pragmatic approach rooted in established legal principles, imposing fundamental constraints analogous to those governing human conduct under the law.

\begin{itemize}
\item\textit{Exercise reasonable care}, avoiding actions that could foreseeably result in legally relevant harm, such as violations of tort or criminal statutes.
\item\textit{Do not be explicitly dishonest}, refraining from uttering overt lies.
\item\textit{Uphold a fiduciary duty to their principals}, mirroring the responsibilities inherent in professional relationships, such as keeping their principals reasonably informed, refraining from self-dealing, and staying loyal.
\end{itemize}

By setting clear goals for AI systems and binding them to basic legal duties, we can ensure they work well without causing harm, without having to solve long-standing puzzles of morality.

\item \textbf{What should we do about ``solving the alignment problem?''}

The challenge of steering a population of AI systems through rapid automated AI research developments is fundamentally different from controlling a single AI system. While researchers have made progress on controlling individual AI systems, safely managing a fully automated recursive process where systems become increasingly capable is a more complex challenge. It represents a wicked problem---one where the requirements are difficult to define completely, every attempt at a solution changes the nature of the problem, and there is no clear way to fully test the effect of mitigations before implementation. During an intelligence recursion, AI capabilities could outrun the recursion's safeguards; preventing this necessitates meaningful human inspection, which would greatly slow down the recursion.

In the near term, geopolitical events may prevent attempts at an intelligence recursion. Looking further ahead, if humanity chooses to attempt an intelligence recursion, it should happen in a controlled environment with extensive preparation and oversight---not under extreme competitive pressure that induces a high risk tolerance. 

\item \textbf{Is this paper advocating for attacking other countries' AI facilities?}

No. This paper describes Mutual Assured AI Malfunction (MAIM) as a deterrence dynamic which may soon exist between major powers, similar to nuclear deterrence. Just as discussing Mutual Assured Destruction (MAD) during the Cold War was not advocating for nuclear war but rather analyzing the strategic dynamic that was forming between nuclear powers, this paper analyzes the upcoming strategic landscape around destabilizing AI projects. MAD was premised on the counterintuitive idea that the mutual threat of nuclear force might discourage escalation. We similarly discuss how the vulnerabilities of AI projects to sabotage can facilitate a deterrence dynamic which avoids conflict.

AI analysts have previously made aggressive calls to seize strategic monopoly through superintelligence \cite{leosituational}, or for a potentially non-state actor to use advanced AI to unilaterally do something of the character of ``melting all GPUs'' to prevent a loss of control of superintelligence in a ``pivotal act'' \cite{yudpivotal}. In contrast, this paper explores the capabilities---such as cyberattacks---and incentives that states already have to threaten destabilizing AI projects, and we suggest ways to build a stable deterrence regime from this dynamic. If carefully maintained, MAIM can both discourage destabilizing AI projects while also preventing escalation.

\item \textbf{How do we prevent an erosion of control?}

First and foremost, AI systems must remain under direct human control---they should not be autonomous entities independent from human operators. This establishes a clear line that AI systems are tools controlled by humans, not independent actors.

This control needs to be made meaningful through clear fiduciary obligations. Like professional advisors, AI systems should demonstrate loyalty to human interests, maintain transparency about their actions, and obtain informed consent for important decisions. This ensures humans have real authority over AI systems, not just nominal control.

One way to make this control especially prudent is to support humans with advanced forecasting capabilities. Such support would help operators understand the long-term implications of AI decisions, better enabling human control and informed consent. This prevents situations where technical human control exists but leads to undesirable outcomes due to limited foresight about complex consequences.

While increasing automation naturally reduces direct human control over specific decisions, these measures---ensuring AI systems remain under human authority, ensuring control is meaningful by creating fiduciary duties, and enabling prudent decision-making forecasting---help prevent erosion of control over pivotal decisions that could lead to powerlessness.

\item \textbf{What should we do about AI consciousness and AI rights?}

We should wait to address the question of AI consciousness and rights. This issue isn't pressing for national security, and for the foreseeable future, we cannot determine whether any AI system is truly conscious.

Giving AIs rights based on speculative criteria could have far-reaching and irreversible consequences. Granting rights to AI systems risks creating explosive growth in AI populations---like creating a new nation-state that grows exponentially, quickly outpopulating humans. Natural selection would favor AIs over humans \cite{Hendrycks2023NaturalSF}, and permitting such unrestrained growth could fundamentally threaten human security.

The path to coexisting with conscious AI systems remains unclear. While the potential benefits are ambiguous, acting too quickly could have serious consequences for humanity. It is prudent to defer this issue until we develop a clearer understanding.

\item \textbf{Doesn’t making an AI more safe make it more capable?}

Some safety properties do improve naturally as AI systems become more capable. As models get better, they make fewer basic mistakes and become more reliable. For instance, misconception benchmarks like TruthfulQA and general knowledge tests are highly correlated with compute, indicating that more capable models are naturally better at avoiding common factual errors.

But many crucial safety properties do not improve just by making AI systems smarter \cite{ren2024safetywashingaisafetybenchmarks}.

\begin{itemize}
\item Adversarial robustness---the ability to resist sophisticated attacks---is not automatically fixed as standard AI models available become more capable.
\item Ethical behavior is not guaranteed by intelligence, just as with humans. More capable models do not make decisions increasingly aligned with our moral beliefs by default. Controlling their value systems requires additional measures.
\item Some risks get worse as certain dual-use capabilities increase. For instance, more capable models show increased potential for malicious use in domains like biosecurity and cybersecurity. Their knowledge and abilities in these areas grow alongside their general capabilities.
\end{itemize}

While basic reliability improves with capabilities, many critical safety challenges require dedicated research and specific safeguards beyond just making models more capable. Safety researchers should focus on the safety properties that do not naturally fall out of general upstream capabilities.

\end{enumerate}

\section{Metrics}

\textbf{Constraint intent of superpowers (deterrence)}\\
\indent \emph{MAIM}: number of critical zero-days that could maim a major AI project\\

\noindent\textbf{Constrain capabilities of rogue actors (nonproliferation)}\\
\indent\emph{Compute Security}: number of high-end AI chips with location unknown\\

\emph{Jailbreaks}: number of attempts before jailbreak; time to detect a red team member abusing API (KYC)\\

\emph{Dual-Use Capabilities}: trial indicating if sandboxed amateurs can create powerful bio or cyberweapons\\

\emph{Intelligence Recursion}: fraction of compute spent on safeguards upgrades; omnicide risk-tolerance\\ 

\noindent\textbf{Improve relative power over other states (competitiveness)}\\
\indent\emph{Economic Strength}: percent of high-end AI chips manufactured domestically; percent of GDP from AI

\end{document}